\documentclass[pdflatex,sn-mathphys-num]{sn-jnl}
\usepackage{graphicx}%
\usepackage{multirow}%
\usepackage{amsmath,amssymb,amsfonts}%
\usepackage{amsthm}%
\usepackage{mathrsfs}%
\usepackage{float}
\usepackage{color}
\usepackage[title]{appendix}%
\usepackage{xcolor}%
\usepackage{textcomp}%
\usepackage{manyfoot}%
\usepackage{booktabs}%
\usepackage{algorithm}%
\usepackage{adjustbox}%
\usepackage{subcaption}%
\usepackage{algorithmicx}%
\usepackage{algpseudocode}%
\usepackage{listings}%
\usepackage{dsfont}%
\usepackage{pgfplots}
\usepackage{subcaption}
\usepackage{tikz} 

\theoremstyle{thmstyleone}%
\newcommand{\iunit}{\mathrm{i}}

\theoremstyle{thmstyletwo}%

\theoremstyle{thmstylethree}%

\begin{document}

\title[Thermalization of coarse-grained variables in the Toda lattice]{Thermalization of coarse-grained variables in the Toda lattice: a numerical study}

\author[1,2]{\fnm{Martina} \sur{Giachello}}\email{martina.giachello@gssi.it}

\author[3,4]{\fnm{Cesare} \sur{Vianello}}\email{cesare.vianello@phd.unipd.it}

\author*[3,4]{\fnm{Giacomo} \sur{Gradenigo}}\email{giacomo.gradenigo@unipd.it}

\affil[1]{\orgname{Gran Sasso Science Institute},
  \orgaddress{\street{Via F. Crispi 7}, \postcode{67100}
    \city{L'Aquila}, \country{Italy}}}

\affil[2]{\orgname{INFN-LNGS}, \orgaddress{\street{Via G. Acitelli
      22}, \postcode{67100} \city{Assergi (AQ)}, \country{Italy}}}

\affil[3]{\orgdiv{Dipartimento di Fisica e Astronomia ``Galileo
    Galilei''}, \orgname{Università di Padova},
  \orgaddress{\street{Via F. Marzolo 8}, \postcode{35131}
    \city{Padova}, \country{Italy}}}

\affil[4]{\orgname{INFN, Sezione di Padova}, \orgaddress{\street{Via
      F. Marzolo 8}, \postcode{35131} \city{Padova}, \country{Italy}}}

\abstract{We present here a numerical study of the Toda lattice
  Hamiltonian dynamics showing how, despite model integrability, it is
  possible to find many macroscopic observables which display standard
  equilibrium behavior. In particular, we study numerically the energy
  time correlation of certain coarse-grained variables, showing that
  it decays with either exponential or stretched exponential
  behaviour, thus indicating good thermalization properties. The
  spirit of our observation is complementary to that of standard
  Fermi-Pasta-Ulam-Tsingou-like numerical experiments, in which the
  breaking of integrability is regarded as the key ingredient for
  relaxation to thermal equilibrium. The present communication
  corroborates, on the contrary, the view that integrability is not an
  obstacle to thermalization, provided that a sufficiently weak
  definition of the latter is accepted. In the Toda lattice we have
  introduced several kinds of coarse-grained variables, constructed
  with the purpose to maintain a thermodynamic relevance also for
  large system sizes, which all show good thermalization properties.}


\maketitle

\newpage
\section{Introduction}\label{sec1}

To celebrate the 70th anniversary of the seminal paper by Fermi, Pasta, Ulam and Tsingou (FPUT) on transient ergodicity breaking in a chain of weakly nonlinear springs~\cite{Fermi1955}, we have chosen to consider the problem of thermalization from another perspective, namely the one originally adopted by Khinchin~\cite{khinchin_book} for classical systems and by Von Neumann~\cite{vonNeumann2010} for quantum systems. Rather than investigating the role played by nonlinearities to achieve thermalization, we try to emphasize how thermalization is a phenomenon intrinsically due to the presence of a large number of degrees of freedom, provided that sufficiently generic observables are considered. This perspective, which has recently reappeared in several works on the foundations of statistical mechanic~\cite{Baldovin_2021,COCCIAGLIA2022127581,Cattaneo25,Baldovin_review}, is complementary to the one usually adopted for nonlinear systems, in which non-integrability and chaoticity of the microscopic dynamics are regarded as the key ingredients to guarantee thermalization. Indeed, taking for granted that the breaking of integrability in the Liouville-Arnold theorem sense is a sufficient condition for a system to reach thermal equilibrium, the central question of the research line initiated by the FPUT paper has always been to characterize how the {\it distance} from integrability relates to the extent of relaxational time scales. The FPUT pioneering numerical investigation gave rise to a fertile field of research which, over the following the decades, produced crucial results in nonlinear dynamics and integrable systems (actually leading to the discovery of some among them). These results range from the discovery of solitons in the Korteweg-de Vries equation, obtained as a continuum limit of the FPUT chain~\cite{Zabusky}, to the recent finding that, at sufficiently low energies, the long time scales observed in certain FPUT experiments are due to the proximity of the Hamiltonian to that of the integrable Toda lattice~\cite{Benettin2013}. The study of transient ergodicity breaking in nonlinear systems is also connected to other very active fields, such as wave turbulence~\cite{Onorato2015}, nonlinear optics~\cite{Pierangeli2018} and disordered systems~\cite{Gradenigo2022}.

Here we take an entirely different perspective: we study a system, the Toda lattice~\cite{TODA_original}, which is integrable in the Liouville-Arnold sense, so that all Lyapunov exponents vanish, there is no chaos, and the dynamics is not mixing~\cite{Henonintegrals}. Nevertheless, as stated by Khinchin's theorem~\cite{khinchin_book} and numerically substantiated in Ref.~\cite{Baldovin_2021}, for \emph{macroscopic} observables, i.e., functions of many almost independent degrees of freedom, the results of equilibrium statistical mechanics apply irrespective of the nature of the microscopic dynamics, that is, regardless of whether it is chaotic or integrable. Let us give a qualitative explanation of why the energy of a Fourier mode of the Toda lattice is expected to show good thermalization properties despite the integrability of the model, as found numerically in Ref.~\cite{Baldovin_2021}. According to Khinchin's theorem, for almost all initial data, dynamical and ensemble averages are expected to coincide for any \emph{sum function}, i.e., a function of the form $F(X)=\sum_{i=1}^M f(x_i)$, with $M \gg 1$ and the $x_i$ almost independent variables. If we regard the Toda actions $\mathcal{I}_i$ as independent variables (which, owing to integrability, holds even in a strict sense), then in the highly nonlinear regime the energy $E_k$ of the Fourier mode $k$ can indeed be regarded as a sum function of these variables, $E_k = \sum_{i=1}^M f_k(\mathcal{I}_i)$, for some function $f_k$. This simple observation rationalizes the results of Ref.~\cite{Baldovin_2021}. However, in the large-$N$ limit, where $N$ is the number of particles in the lattice, the energy of a single Fourier mode remains essentially a microscopic variable, so its relevance for the thermodynamics of the system can still be questioned~\cite{Benettin25-FPUT}. For this reason, in the present paper we extend the studies of Refs.~\cite{Baldovin_2021,COCCIAGLIA2022127581,Cattaneo25} to the relaxation properties of \emph{groups} of microscopic variables comprising a finite fraction of the total number of degrees of freedom. These groups, discussed in detail throughout the paper, act as coarse-grained variables that retain their thermodynamic relevance in the large-$N$ limit. To this purpose we also recall some recent results on the standard FPUT problem, which likewise supports the view that integrability and the validity of equilibrium ensembles are not in contradiction: the equilibrium fluctuations of the high-energy dynamics of FPUT models with quartic nonlinear interactions are well reproduced by an effective integrable Hamiltonian, showing that such models are, in a statistical sense, equivalent to an integrable one even in a regime of strong chaoticity~\cite{Ponno2026}.

The paper is organized as follows. In Sec.~\ref{sec:model} we introduce the model, the integrable Toda lattice, and define the different sets of \emph{microscopic} variables considered: Fourier modes, random modes, and quasi-particle modes. Sec.~\ref{sec:observables} presents the coarse-grained observables, namely the energies of groups of the variables introduced in Sec.~\ref{sec:model}, whose time autocorrelation functions are the object of our study. It also describes the numerical protocol: canonical Monte Carlo sampling to prepare the initial conditions, followed by microcanonical Hamiltonian dynamics to sample the equilibrium fluctuations. In Sec.~\ref{sec:autocorrelationcoarsegrained} we present and discuss the numerical results, which all point to the same evidence: exponential relaxation to equilibrium. Finally, Sec.~\ref{sec:conclusions} contains our conclusions.

\section{The model}\label{sec:model}

The Toda lattice is a one-dimensional chain of $N$ classical particles interacting via a nearest-neighbour exponential
potential~\cite{TODA_original}. By denoting with $q_i$ the displacement of the $i$-th particle from its equilibrium position, with $p_i$ its conjugate momentum, where $i=1,\ldots,N$, and by $V(x)$ the nonlinear potential
\begin{equation}
    V(x) = e^{-x} + x - 1,
    \label{eq:toda-potential}
\end{equation}
we have that the dynamics of the Toda model is the one generated by the following Hamiltonian
\begin{align}
  \mathcal{H}(q,p)
  = \sum_{i=1}^N \frac{p_i^2}{2}
  + \sum_{i=0}^{N} V(q_{i+1}-q_i).
  \label{eq:HToda-model}
\end{align}
In the present work we consider the case of fixed (Dirichlet) boundary conditions $q_0 = q_{N+1} = 0$. The Toda lattice has $N$ independent integrals of motion $\mathcal{I}_1,\ldots,\mathcal{I}_N$ in involution, satisfying $\{\mathcal{I}_j,\mathcal{H}\}=0$ and $\{\mathcal{I}_j,\mathcal{I}_\ell\}=0$ for all $j,\ell=1,\dots,N$~\cite{Henonintegrals}. The model is therefore integrable in the sense of the Liouville-Arnold theorem: the phase-space dynamics takes place on $N$-dimensional invariant tori and trajectories are quasi-periodic \cite{Arnold}. 

\subsection{Fourier Modes}

Despite model's integrability, it was recently shown in Ref.~\cite{Baldovin_2021} that thermalization can be observed even in the Toda lattice, provided that suitable observables are considered, in particular collective variables weakly
correlated with the Toda actions. The main result of~\cite{Baldovin_2021} has been to show that, at high non-perturbative energies, this is precisely the case for Fourier modes. For this reason we will consider also here Fourier modes, which are defined as
\begin{subequations}
\label{eq:FourierModes}
\begin{align}
    Q_k &= \sqrt{\frac{2}{N+1}}
    \sum_{i=1}^N q_i
        \sin\!\left(\frac{\pi i k}{N+1}\right),\\
    P_k &= \sqrt{\frac{2}{N+1}}
    \sum_{i=1}^N p_i
        \sin\!\left(\frac{\pi i k}{N+1}\right),
        \label{eq:Fourier-definition}
\end{align}
\end{subequations}
with $k=1,\ldots,N$. By performing the canonical change of variables $(q_i,p_i)\rightarrow{(Q_k,P_k)}$, the Toda Hamiltonian can be rewritten as
\begin{align}
  \mathcal{H}(Q,P)
  = \frac{1}{2}\sum_{k=1}^N
    (P_k^2 + \omega_k^2 Q_k^2)
    + \sum_{n\ge 3}
      \sum_{k_1,\ldots,k_n}
      \alpha_{k_1\ldots k_n}
      Q_{k_1}\cdots Q_{k_n},
  \label{eq:toda-ham-fourier}
\end{align}
where the coefficients $\alpha_{k_1\ldots k_n}$ parametrize the nonlinear couplings between Fourier modes and the frequencies $\omega_k$ satisfy the dispersion relation
\begin{equation}
    \omega_k = 2 \sin\!\left(\frac{\pi k}{2(N+1)}\right).
    \label{eq:dispersion}
\end{equation}
For our purposes, it is convenient to introduce the \emph{complex Fourier modes}
\begin{align}
    z_k = \frac{P_k+\iunit \omega_kQ_k}{\sqrt{2\omega_k}},\qquad z^*_k = \frac{P_k- \iunit\omega_kQ_k}{\sqrt{2\omega_k}},
\end{align}
satisfying $\{z_k,z^*_q\}=\iunit \delta_{kq}$. In Ref.~\cite{Baldovin_2021} it was shown that the energy of each Fourier mode,
\begin{equation}
    \mathcal{E}^{(\mathrm F)}_k(t)
    = \frac{1}{2}\left[P_k^2(t) + \omega_k^2 Q_k^2(t)\right] = \omega_k|z_k(t)|^2,
    \label{eq:fourier-modes}
\end{equation}
which is not conserved by the Toda dynamics, thermalizes quickly, relaxing to the equipartition value.

\subsection{Random Modes}

The same conclusion on the thermalization of Fourier modes in the Toda model proposed in~\cite{Baldovin_2021}, was drawn in Ref.~\cite{COCCIAGLIA2022127581} for the energies of \emph{random modes} $z_k(\boldsymbol\theta)$ in the classical harmonic chain. These modes are obtained from the complex Fourier modes through a random unitary transformation $M(\boldsymbol{\theta}) \in U(N)$, drawn from the Haar measure, where $\boldsymbol{\theta}$ collectively denotes the $N^2$ real parameters (angles and phases) identifying the realization:
\begin{equation}
    z_k(\boldsymbol{\theta})
    = \sum_{q=1}^N M_{kq}(\boldsymbol{\theta}) z_q.
    \label{eq:random-modes}
\end{equation}
Being unitary, the transformation preserves the Poisson brackets: $\{z_k(\boldsymbol{\theta}), z_q^*(\boldsymbol{\theta})\} = \iunit \delta_{kq}$. In terms of the random modes, the quadratic part of the Hamiltonian of Eq.~\eqref{eq:toda-ham-fourier} reads
\begin{align}
    \mathcal{H}_2(z(\boldsymbol{\theta}),z^*(\boldsymbol{\theta}))
    &= \sum_{q,p=1}^N c_{qp}(\boldsymbol{\theta})z_q^*(\boldsymbol{\theta})z_p(\boldsymbol{\theta}),
    \label{eq:H-random-toda}
\end{align}
where 
\begin{equation}
    c_{qp}(\boldsymbol{\theta})
    = \sum_{k=1}^N \omega_k M_{qk}(\boldsymbol{\theta})M^*_{pk}(\boldsymbol{\theta}).
\end{equation}
Thus, in terms of the random modes, even the quadratic part of the Hamiltonian is not diagonal. This allows energy to be exchanged among the random modes, leading to the results of Ref.~\cite{COCCIAGLIA2022127581}. From the diagonal terms of Eq.~\eqref{eq:H-random-toda}, we define the energy of each random mode as
\begin{equation}
    \mathcal{E}^{(\mathrm R)}_k(\boldsymbol{\theta},t) = c_{kk}(\boldsymbol\theta)|z_k(\boldsymbol\theta,t)|^2.
\end{equation}

\subsection{Quasi-particle modes}\label{subsec:into_quasi_particle}

In the present study we also consider a third set of variables which, to the best of our knowledge, has not been used before in thermalization studies of FPUT, Toda or related models. These variables are obtained by applying the inverse sine transform to the complex Fourier modes:
\begin{equation}
  \zeta_i =
  \sqrt{\frac{2}{N+1}}
  \sum_{k=1}^N
  \sin\!\left(\frac{\pi i k}{N+1}\right)z_k.
  \label{eq:zeta_def_model}
\end{equation}
Since the transformation is orthogonal, the new variables also satisfy $\{\zeta_i,\zeta_j^*\}={\rm i}\delta_{ij}$. Then, the quadratic part of the Hamiltonian \eqref{eq:toda-ham-fourier} reads
\begin{equation}
  \mathcal H_2(\zeta, \zeta^*)=\sum_{i,j=1}^N a_{ij} \zeta_i^* \zeta_j,
  \label{eq:zetai-Ham}
\end{equation}
with
\begin{equation}
  a_{ij} =
  \frac{2}{N+1}\sum_{k=1}^N
  \omega_k
  \sin\!\left(\frac{\pi i k}{N+1}\right)
  \sin\!\left(\frac{\pi j k}{N+1}\right).
\end{equation}
We refer to the variables $\zeta_i$ as \emph{quasi-particle modes}. The reason is that, although each $\zeta_i$ depends on all the positions $q_j$ and momenta $p_j$, its dependence on the positions is well resolved in space, as shown in Appendix~\ref{app:zeta_loc}. As for the random modes, the quadratic part of the Hamiltonian is already non-diagonal in these variables. From the diagonal terms of Eq.~\eqref{eq:zetai-Ham}, we define the energy of each quasi-particle mode as
\begin{equation}
  \mathcal E_i^{(\text{QP})}(t) = a_{ii}|\zeta_i(t)|^2.
  \label{eq:zeta_energy_model}
\end{equation}

Having defined all the microscopic variables that will be considered in our study, in the next section we use them to construct \emph{macroscopic} observables involving a finite fraction of the $N$ degrees of freedom.

\section{Correlation functions and numerical protocol}
\label{sec:observables}

Previous works on the thermalization of integrable models focused on the relaxation of single modes from atypical initial conditions, namely Fourier modes in the Toda lattice~\cite{Baldovin_2021} and random modes in the harmonic chain~\cite{COCCIAGLIA2022127581,Cattaneo25}. The problem is that the relaxation of individual degrees of freedom, whether localized or not, might be rightfully considered a poor evidence to conclude that the system as a whole is thermal~\cite{Benettin25-FPUT}. It is for this reason that we have decided to focus the analysis of the present paper on variables relevant even macroscopically. In particular, taking inspiration from~\cite{CGCP07} for the choice of the observable and initial conditions, we consider the time autocorrelation functions of energy along Hamiltonian trajectories starting from typical initial conditions draw from the Boltzmann distribution. But then, rather than studying how decorrelates the energy of a single degree of freedom, as done in~\cite{CGCP07}, we consider here the energy of \emph{groups} (or \emph{packets}) of many degrees of freedom, since the collective behavior of such groups is expected to be more relevant for thermodynamics than the relaxation of individual modes. In particular, we study the time autocorrelation functions of the energies of:
\begin{enumerate}
\item Packets of Fourier modes, defined in Eq.~\eqref{eq:Fourier-definition}, either \emph{ordered} or \emph{randomly chosen}, as defined below;
\item Packets of random modes, defined in Eq.~\eqref{eq:random-modes}, for which there is no distinction between ordered and randomly chosen; 
\item Packets of quasi-particle modes, defined in Eq.~\eqref{eq:zeta_def_model}, again either ordered or randomly chosen.
\end{enumerate}
We recall that the energies of the individual modes are
\begin{subequations}
\begin{align}
\text{Fourier Modes}\qquad &\mathcal{E}^{(\mathrm F)}_k(t)=\omega_k|z_k(t)|^2,\\
\text{Random Modes}\qquad &\mathcal{E}^{(\mathrm R)}_k(\boldsymbol{\theta},t)=c_{kk}(\boldsymbol{\theta})|z_k(\boldsymbol{\theta},t)|^2,\\
\text{Quasi-particle modes}\qquad &\mathcal{E}^{(\text{QP})}_i(t)=a_{ii}|\zeta_i(t)|^2.
\end{align}
\end{subequations}
For Fourier modes, we define the energies of ordered and randomly chosen packets as
\begin{align}
E^{(\mathrm F)}_{\text{ord}}(M,t) &= \frac{1}{M}\sum_{k=1}^M \mathcal{E}^{(\mathrm F)}_k(t)\qquad\text{Ordered packet},\label{eq:e_f_ord}\\ 
E^{(\mathrm F)}_{\text{rand}}(M,t) &= \frac{1}{M}\sum_{\ell=1}^{M} \mathcal{E}^{(\mathrm F)}_{\sigma_\ell}(t)\qquad\textrm{Randomly chosen packet}.
\end{align}
where $\sigma_1,\ldots,\sigma_M$ are the first $M \le N$ entries of a random permutation $\sigma$ of $\{1,\dots,N\}$. An ordered packet thus contains the $M$ lowest-frequency modes, while a random packet contains $M$ modes chosen uniformly at random. Packets of quasi-particle modes are defined in the same way:
\begin{align}
E^{(\text{QP})}_{\mathrm{ord}}(M,t) &= \frac{1}{M}\sum_{i=1}^M \mathcal{E}^{(\text{QP})}_i(t), \\ 
E^{(\text{QP})}_{\mathrm{rand}}(M,t)  &= \frac{1}{M}\sum_{\ell=1}^{M} \mathcal{E}^{(\text{QP})}_{\sigma_\ell}(t).
\end{align}
The distinction between ordered and random packets has an analogous meaning for Fourier and quasi-particle modes. Neighboring indices $k$ of the Fourier modes $z_k$ correspond to neighboring frequencies $\omega_k$, i.e., to similar time scales, whereas neighboring indices $i$ of the quasi-particle modes $\zeta_i$ correspond to excitations localized at neighboring positions in space (see Appendix~\ref{app:zeta_loc}). Choosing the indices at random instead mixes different time scales in the first case, and excitations localized at distant positions in the second. For random modes, by contrast, the index $k$ labeling $z_k(\boldsymbol{\theta})$ carries no information on either frequency or spatial position, so ordered and random packets are statistically equivalent. In this case we therefore simply consider
\begin{equation}
E_{\boldsymbol{\theta}}^{(\text{R})}(M,t) = \frac{1}{M}\sum_{k=1}^{M} \mathcal{E}^{(\text{R})}_{k}(\boldsymbol{\theta},t). \label{eq:e_rand}
\end{equation}
Throughout, we assume the equivalence between ensemble and time averages, and compute all averages along trajectories. Accordingly, for a generic packet energy $E(M,t)$ defined by Eqs.~\eqref{eq:e_f_ord}-\eqref{eq:e_rand}, the expectation value is computed as 
\begin{subequations}
\label{eq:time-averages}
\begin{align}
\langle E(M)\rangle &= \frac{1}{\tau} \int_0^\tau ds\,E(M,s),
%\\ \langle E(M,t) E(M,0)\rangle &= \frac{1}{\tau} \int_0^\tau ds\,E(M,s+t)\,E(M,s),
\end{align}
\end{subequations}
where $\tau$ is the total integration time, while in the two-point correlation function $\langle E(M,t) E(M,0)\rangle$ the angular parentheses denote a sliding time-window average. We can now introduce the main observables. These are the normalized time autocorrelation function of the energies of ordered and randomly chosen packets of Fourier modes,
\begin{align}
C^{(\text{F})}_{\text{ord}}(M,t) &= \frac{\left\langle E^{(\text{F})}_{\text{ord}}(M,t)~E^{(\text{F})}_{\text{ord}}(M,0)\right\rangle
    - \left\langle E^{(\text{F})}_{\text{ord}}(M)\right\rangle^2
  }{\left\langle E^{(\text{F})}_{\text{ord}}(M)^2\right\rangle
    -\left\langle E^{(\text{F})}_{\text{ord}}(M)\right\rangle^2}, \label{eq:CF_ord} \\
   C^{(\text{F})}_{\text{rand}}(M,t)&= \frac{\left\langle E^{(\text{F})}_{\text{rand}}(M,t)~E^{(\text{F})}_{\text{rand}}(M,0)\right\rangle
    - \left\langle E^{(\text{F})}_{\text{rand}}(M)\right\rangle^2
  }{\left\langle E^{(\text{F})}_{\text{rand}}(M)^2\right\rangle
    -\left\langle E^{(\text{F})}_{\text{rand}}(M)\right\rangle^2}.
    \label{eq:CF_dis}
\end{align}
Analogously, for ordered and randomly chosen packets of quasi-particle modes, we define
\begin{align}
C^{(\text{QP})}_{\text{ord}}(M,t) &= \frac{\left\langle E^{(\text{QP})}_{\text{ord}}(M,t)~E^{(\text{QP})}_{\text{ord}}(M,0)\right\rangle
    - \left\langle E^{(\text{QP})}_{\text{ord}}(M)\right\rangle^2
  }{\left\langle E^{(\text{QP})}_{\text{ord}}(M)^2\right\rangle
    -\left\langle E^{(\text{QP})}_{\text{ord}}(M)\right\rangle^2}, \label{eq:CQP_ord} \\
   C^{(\text{QP})}_{\text{rand}}(M,t)&= \frac{\left\langle E^{(\text{QP})}_{\text{rand}}(M,t)~E^{(\text{QP})}_{\text{rand}}(M,0)\right\rangle
    - \left\langle E^{(\text{QP})}_{\text{rand}}(M)\right\rangle^2
  }{\left\langle E^{(\text{QP})}_{\text{rand}}(M)^2\right\rangle
    -\left\langle E^{(\text{QP})}_{\text{rand}}(M)\right\rangle^2},
    \label{eq:QP_dis}
\end{align}
and, finally, for packets of random modes
\begin{align}
C_{\boldsymbol{\theta}}^{(\text{R})}(M,t) &= \frac{\left\langle E_{\boldsymbol{\theta}}^{(\text{R})}(M,t)~E_{\boldsymbol{\theta}}^{(\text{R})}(M,0)\right\rangle
    - \left\langle E_{\boldsymbol{\theta}}^{(\text{R})}(M)\right\rangle^2
  }{\left\langle E_{\boldsymbol{\theta}}^{(\text{R})}(M)^2\right\rangle
    -\left\langle E_{\boldsymbol{\theta}}^{(\text{R})}(M)\right\rangle^2}. \label{eq:CR} 
\end{align}

We now specify the numerical integration scheme and the choice of initial conditions. We integrate Hamilton's equations for the particle variables,
\begin{subequations}
\begin{align}
\dot{q}_i = \frac{\partial\mathcal{H}}{\partial p_i} &= p_i,\\
\dot{p}_i = -\frac{\partial\mathcal{H}}{\partial q_i} &= e^{-(q_i-q_{i-1})} - e^{-(q_{i+1}-q_i)},
\end{align}
\end{subequations}
for $i=1,\dots,N$ with $q_0=q_{N+1}=0$, using the same symplectic leap-frog integrator as in Ref.~\cite{Baldovin_2021}, with time step $\delta t = 0.01$. In all the numerical simulations discussed in this paper we have performed Hamiltonian dynamics starting from initial configurations drawn from the equilibrium canonical measure at temperature $T=3$. This means that, before running the Hamiltonian dynamics, we have thermalized the positional degrees of freedom using a Monte Carlo algorithm built on the weight $\exp[-\beta \sum_{i=0}^{N} V(q_{i+1}-q_i)]$, while the momenta are drawn independently from a Gaussian distribution with zero mean and variance equal to $T$. Details on the preparation of the initial conditions are given in Appendix~\ref{app:mc_ic}. Once the initial configuration is drawn from equilibrium, the system is evolved under the Hamiltonian dynamics and the correlation functions are computed as time averages along a single trajectory.\\

\section{Coarse-grained variables relaxation}
\label{sec:autocorrelationcoarsegrained}
In this section we present the numerical evidence supporting the main conclusion of this paper: in the Toda lattice there exist coarse-grained variables of thermodynamic relevance that thermalize despite the integrability of the model. We show that the qualitative details of the time autocorrelation functions depend on the observable considered; for instance, their decay may be characterized by one or by several time scales. A single robust feature, however, emerges for all observables: their time autocorrelation functions decay exponentially. This supports the view that thermalization also takes place in integrable systems~\cite{Baldovin_2021,COCCIAGLIA2022127581,Cattaneo25,Baldovin_review}, provided that appropriate observables are considered. Indeed, all the coarse-grained variables considered here closely match the definition of sum function introduced by Khinchin in his proof of the ergodic theorem for such functions~\cite{khinchin_book}. The following subsections present the numerical results for the different coarse-grained variables: packets of Fourier modes in Sec.~\ref{subsec:fourier-modes}, packets of
random modes in Sec.~\ref{subsec:random-modes}, and packets of
quasi-particle modes in Sec.~\ref{subsec:quasiparticle-modes}.
\begin{figure}[t!]
\centering
\begin{subfigure}[t]{0.49\textwidth}
    \centering
    \includegraphics[width=\textwidth]{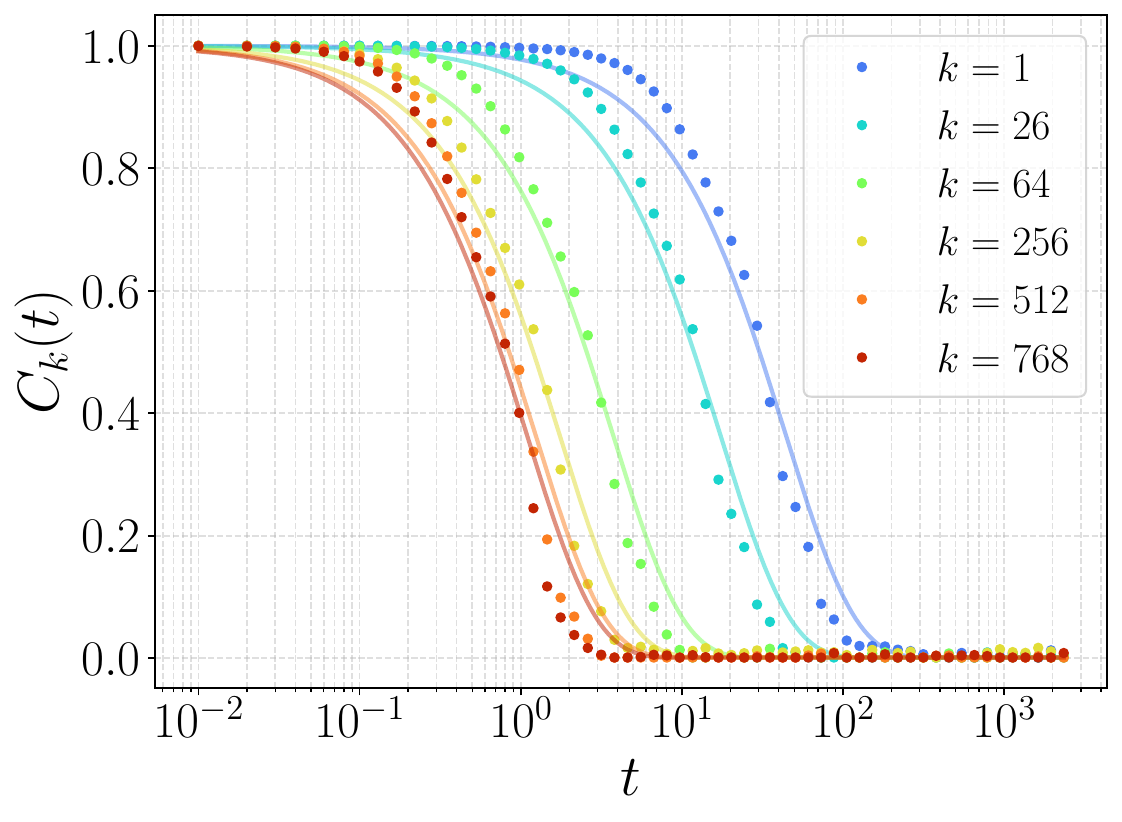}
    \caption{}
    \label{fig:fourier_singlemode_fits}
\end{subfigure}
\hfill
\begin{subfigure}[t]{0.49\textwidth}
    \centering
    \includegraphics[width=\textwidth]{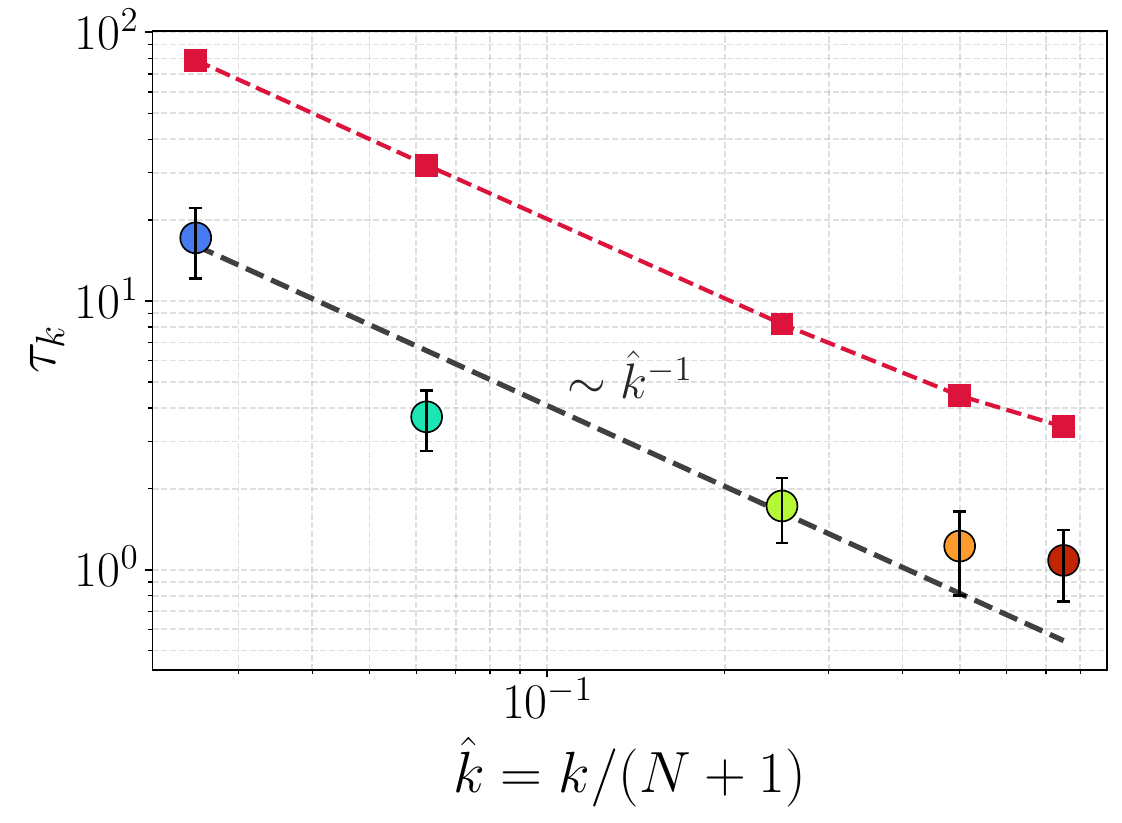}
    \caption{}
    \label{fig:fourier_singlemode_tau}
\end{subfigure}
\caption{Time-correlation functions of single Fourier modes and their characteristic times. Data obtained for a Toda lattice with $N=1023$ degrees of freedom and fixed boundary conditions. Initial data for the Hamiltonian dynamics are drawn from equilibrium Boltzmann distribution at temperature $T=3$ (average energy per degree of freedom $\langle E\rangle/N = 2.73$). The time step of Hamiltonian dynamics (leap-from symplectic integrator) is $\delta\tau=0.01$. {\it \underline{Panel} a)}: Fourier mode autocorrelation $C_k(t)$, see definition in Eq.~\eqref{eq:Ck-single}, for wavenumber $k = 1, 26, 64, 256, 512, 768$. Bullets represents numerical data while continuous lines are exponential fits with $C_k(t) = e^{-t/\tau_k}$, where $\tau_k$ is the characteristic thermalization time for each individual mode. {\it \underline{Panel} b)}: (Blue) Circles are characteristic thermalization times $\tau_k$ as obtained from the exponential fits in panel {\it a)} plotted as a function of the normalized wavenumber $\hat{k} = k/(N+1)$, (red) squares are the oscillation periods $\mathcal{T}_k$ of each Fourier mode, as obtained from the dispersion relation $\mathcal{T}_k=2\pi/\omega_k$, with $\omega_k = 2\sin[k\pi/2(N+1)]$, so that, when $k\ll N$, we have $\omega_k \sim \hat{k} = k/N \rightarrow \mathcal{T}_k \sim \hat{k}^{-1}$, as emphasized by the (red) dotted line.}
\label{fig:fourier_singlemode_autocorr}
\end{figure}

\subsection{Fourier modes packets}\label{subsec:fourier-modes}

The main result concerning packets of Fourier modes is that their relaxation time scale depends on how the packet is built. To gain better insight into the behavior of Fourier mode packets, it is useful first to recall the behavior of the autocorrelation of the harmonic energy of single Fourier modes, already discussed in Ref.~\cite{Baldovin_2021}, and to analyze it in somewhat more detail. Fig.~\ref{fig:fourier_singlemode_autocorr} shows the decay of the correlation functions
\begin{align}
C^{(\rm F)}_k(t) = \frac{\left\langle \mathcal E^{(\rm F)}_k(t)\mathcal E^{(\rm F)}_k(0)\right\rangle - \left\langle \mathcal E^{(\rm F)}_k \right\rangle^2}{\left\langle \bigl[\mathcal E^{(\rm F)}_k\bigr]^2 \right\rangle - \left\langle \mathcal E^{(\rm F)}_k \right\rangle^2}.
\label{eq:Ck-single}
\end{align}
Two main observations can be made on the data shown in Fig.~\ref{fig:fourier_singlemode_fits}. First, all correlation functions decay exponentially, irrespective of the mode index $k$. Second, the characteristic time scale of the exponential decay, which we call $\tau_k$, depends on $k$ in the same way as the characteristic period of the mode, $\mathcal{T}_k = 2\pi/\omega_k$, where $\omega_k$ is the frequency given by the dispersion relation~\eqref{eq:dispersion}. The comparison between $\tau_k$ and $\mathcal{T}_k$ as functions of $k$ is shown in  Fig.~\ref{fig:fourier_singlemode_tau}. Recall that, for $\hat{k} = k/N \ll 1$, the dispersion relation gives $\omega_k \sim \hat{k}$, so that $\mathcal{T}_k \sim \hat{k}^{-1}$; approximately the same behavior is found for $\tau_k$, i.e., $\tau_k \sim \hat{k}^{-1}$.

\begin{figure}[p]
\centering
\begin{subfigure}[t]{0.49\textwidth}
    \centering
    \includegraphics[width=\textwidth]{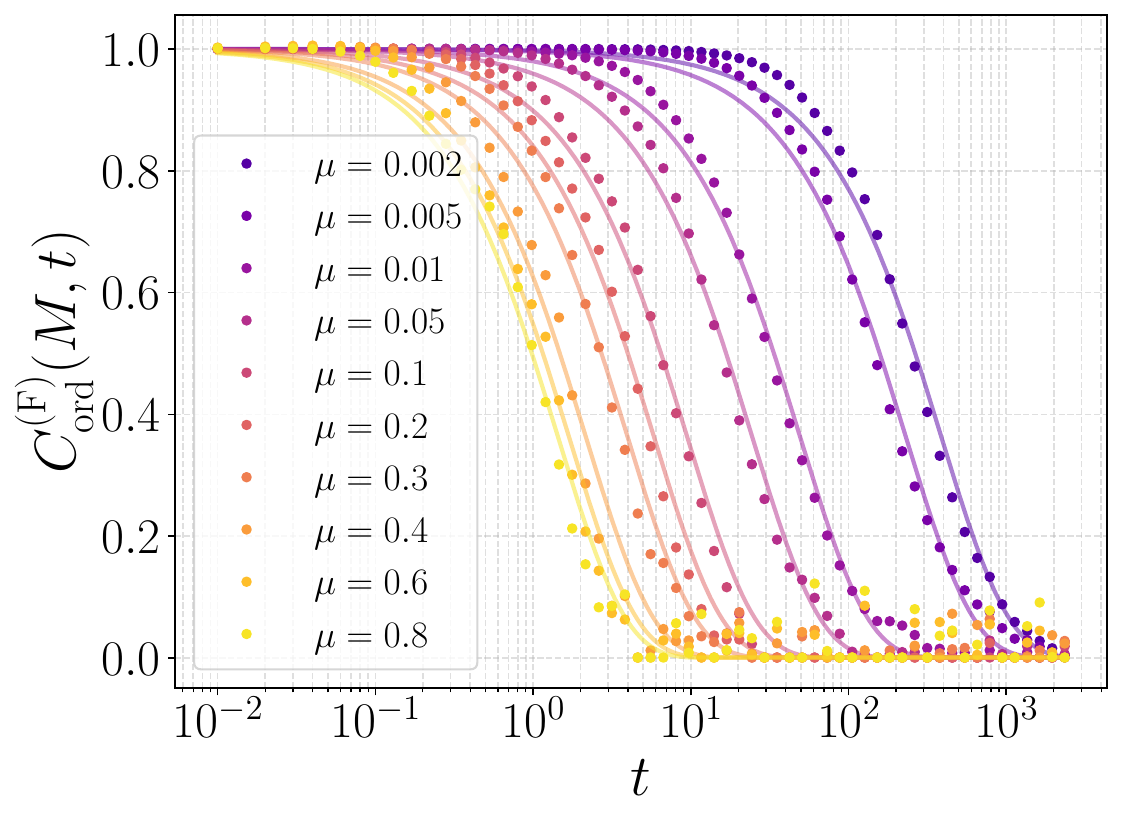}
    \caption{}
    \label{fig:autocorrelation-ordered-fourier-fits}
\end{subfigure}
\hfill
\begin{subfigure}[t]{0.49\textwidth}
    \centering
    \includegraphics[width=\textwidth]{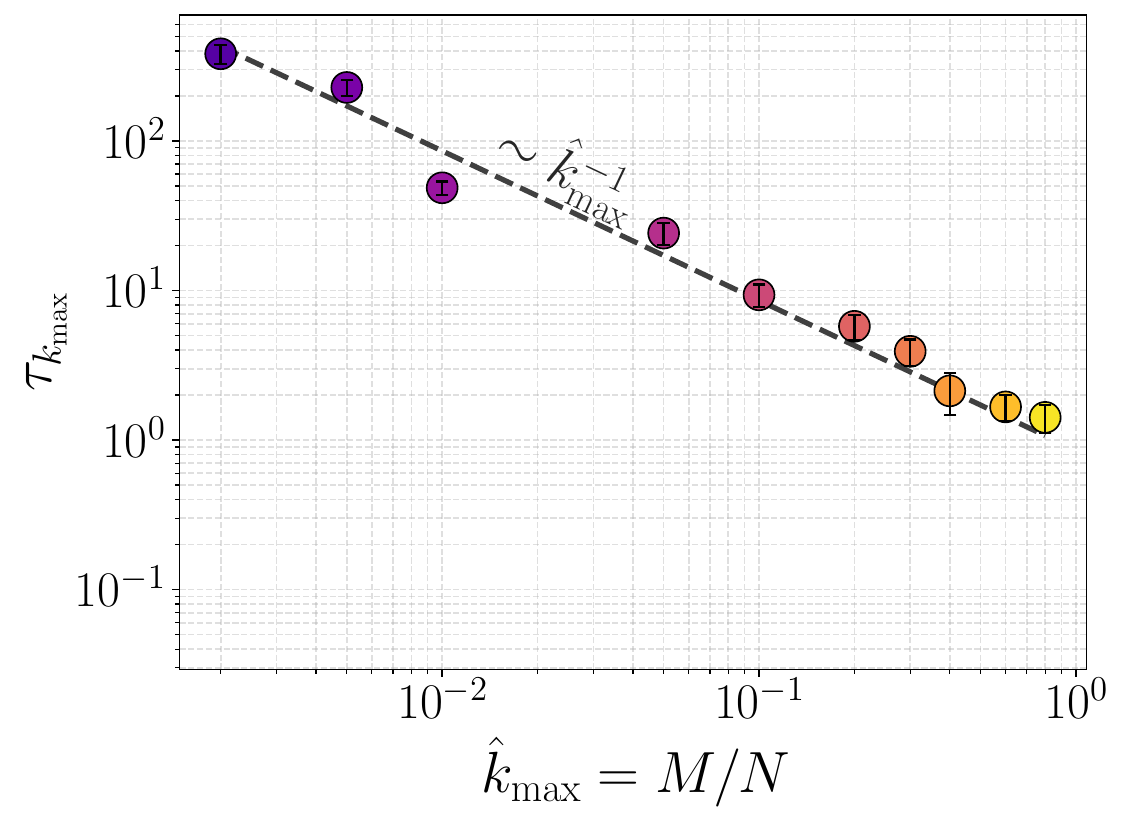}
    \caption{}
    \label{fig:autocorrelation-ordered-fourier-tau}
\end{subfigure}
\hfill
\begin{subfigure}[t]{0.49\textwidth}
    \centering
    \includegraphics[width=\textwidth]{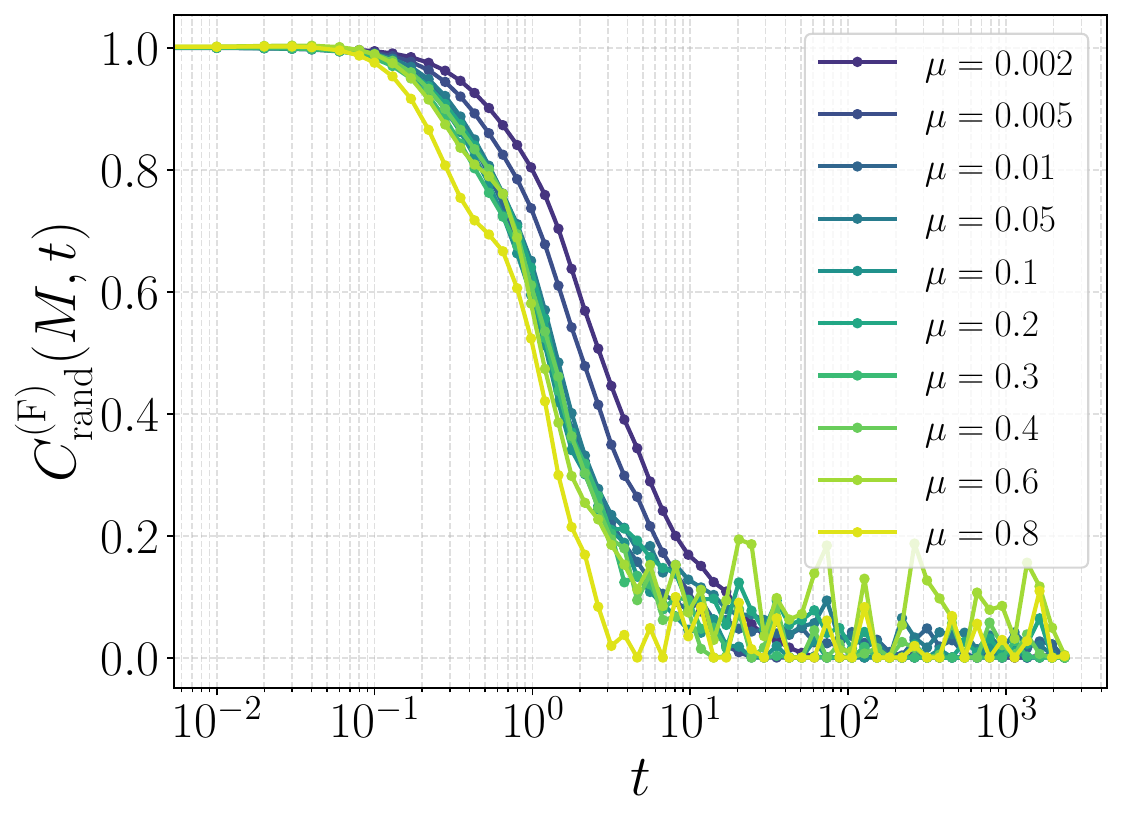}
    \caption{}
    \label{fig:Ebar_random}
\end{subfigure}
\caption{Time-correlation functions of packets of Fourier modes and the characteristic decay times of ordered packets. Data obtained for a Toda lattice
  with $N=1023$ degrees of freedom and fixed boundary
  conditions. Initial data for the Hamiltonian dynamics are drawn from
  equilibrium Boltzmann distribution at temperature $T=3$ (average
  energy per degree of freedom $\langle E\rangle/N = 2.73$). The time step
  of Hamiltonian dynamics (leap-from symplectic integrator) is
  $\delta\tau=0.01$. {\it \underline{Panel} a)}: Bullets are numerical
  data for normalized autocorrelation $C^{(\mathrm
    F)}_{\mathrm{ord}}(M,t)$ (see definition in Eq.~\eqref{eq:CF_ord})
  of Fourier modes ordered packets for size ratio $\mu=M/N$ from
  $\mu=0.002$ to $\mu=0.8$, continuos line are the exponential
  fits. In the case of ordered packets we have that $M$ corresponds to
  the largest wave-number of the packets, so that we can write
  $k_{\text{max}}=M$ (and, correspondingly,
  $\hat{k}_{\text{max}}=M/N$). {\it \underline{Panel} b)}
  Characteristic thermalization times $\tau_{k_{\text{max}}}$ for the
  ordered packets extracted from the exponential fits in panel {\it
    a)}, dotted line emphasizes the power-law scaling
  $\tau_{k_{\text{max}}} \sim \hat{k}_{\text{max}}^{-1}$.  {\it
    \underline{Panel} c)}: Autocorrelation $C^{(\mathrm
    F)}_{\mathrm{rand}}(M,t)$ of random Fourier packets, see
  definition in Eq.~\eqref{eq:CF_dis}, for packet size fraction
  $\mu=M/N$ from $\mu=0.002$ to $\mu=0.8$.}
\label{fig:autocorrelation-ordered-fourier}
\end{figure}

Let us now focus on packets of modes. For ordered packets, the size $M$ corresponds to the largest wavenumber index in the packet, $M = k_{\text{max}}$, since the packet contains the modes with $k = 1,\dots,M$. Hence, the larger the packet, the larger is the amount of \emph{fast} characteristic times $\tau_k$ contributing to its overall relaxation. Remarkably, if we chose to indicate the characteristic exponential decay time scale $\tau_{\text{ord}}(M)$  of an ordered packet of size $M$ using the symbol $\tau_{k_{\text{max}}}$, that is,
\begin{align}
\tau_{k_{\text{max}}} = 
\tau_{\text{ord}}(M),
\end{align}
we find that the time scales $\tau_{k_{\text{max}}}$ of the exponential decay of ordered packets also scale as $\tau_{k_{\text{max}}} \sim \hat{k}_{\text{max}}^{-1}$. Fig.~\ref{fig:autocorrelation-ordered-fourier-fits} shows the decay of the correlation functions $C^{(\rm F)}_{\textrm{ord}}(M,t)$ for different values of $M$, with superimposed exponential fits, while Fig.~\ref{fig:autocorrelation-ordered-fourier-tau} shows the characteristic times $\tau_{k_{\text{max}}}$ obtained from the fits as a function of $\hat{k}_{\text{max}} = M/N$. Beyond the dependence of the relaxation time on the packet size, the most relevant observation is that the decay is exponential for every size. The same holds for $C^{(\rm F)}_{\textrm{rand}}(M,t)$, i.e., when the Fourier modes in the packet are chosen at random, so that both fast and slow single-mode time scales $\tau_k$ contribute to the packet relaxation for any packet size. Therefore, in this case the overall relaxation time does not depend on the packet size, as is evident from the data shown in Fig.~\ref{fig:Ebar_random}.

\begin{figure}[t!]
\centering
\begin{subfigure}[t]{0.49\textwidth}
    \centering
    \includegraphics[width=\textwidth]{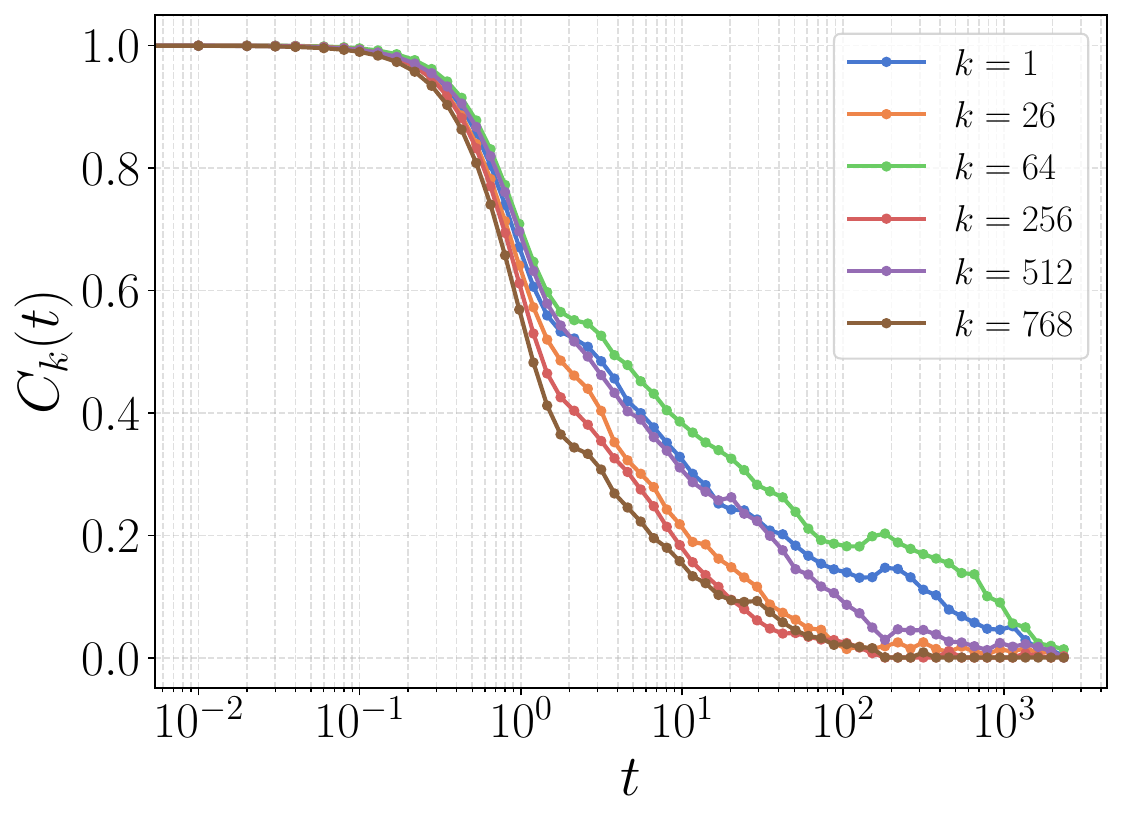}
    \caption{}
    \label{fig:correlations-random-modes-single}
\end{subfigure}
\hfill
\begin{subfigure}[t]{0.49\textwidth}
    \centering
    \includegraphics[width=\textwidth]{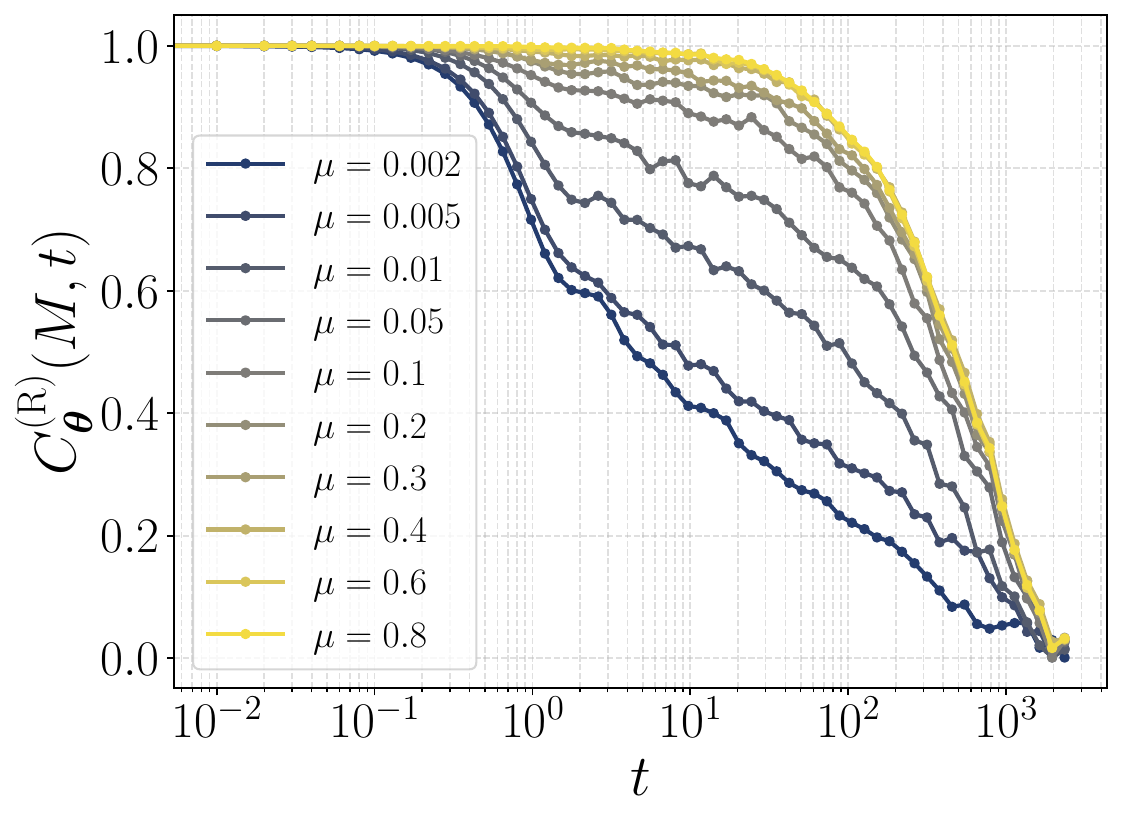}
    \caption{}
    \label{fig:correlations-random-modes}
\end{subfigure}
\hfill
\begin{subfigure}[t]{0.49\textwidth}
    \centering
    \includegraphics[width=\textwidth]{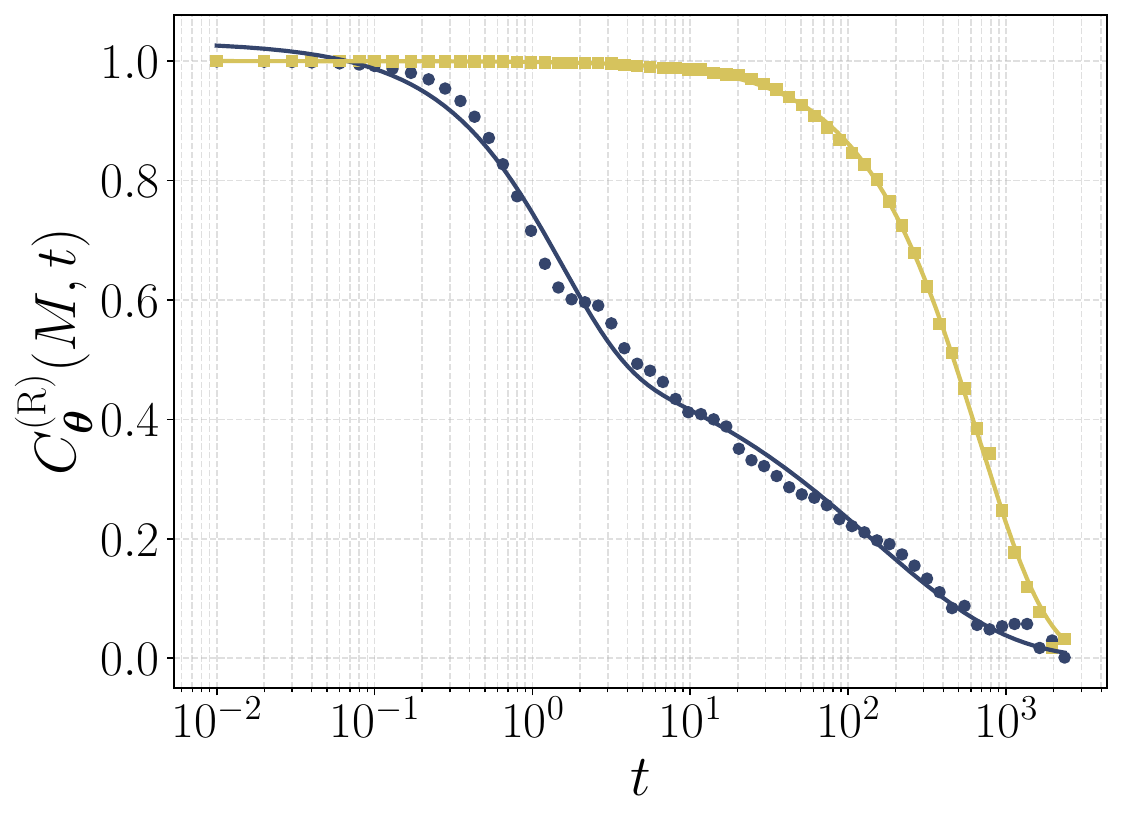}
    \caption{}
    \label{fig:autocorr_Ebar_rotated_fit}
\end{subfigure}
\caption{Time-correlation function of random
  modes, computed for a fixed (quenched) random unitary
  $M(\boldsymbol{\theta})$. Data are obtained for a Toda lattice
  with $N=1023$ degrees of freedom and fixed boundary
  conditions. Initial data for the Hamiltonian dynamics are drawn from
  equilibrium Boltzmann distribution at temperature $T=3$ (average
  energy per degree of freedom $\langle E\rangle/N = 2.73$). The time step
  of Hamiltonian dynamics (leap-from symplectic integrator) is
  $\delta\tau=0.01$. {\it \underline{Panel} a)}: Time-correlation
  $C_k(t)$ of individual random modes energies, as defined in
  Eq.~\eqref{eq:Ck-single}, for different values of the index
  $k$. {\it \underline{Panel} b)}: correlation
  $C_{\boldsymbol{\theta}}^{(\mathrm R)}(M,t)$ for packets of random
  modes, see definition in Eq.~\eqref{eq:CR}, for packets size ratio
  $\mu=M/N$ ranging from $\mu=0.002$ to $\mu=0.8$. {\it \underline{Panel} c)}: autocorrelation
  $C_{\boldsymbol{\theta}}^{(\mathrm{R})}(M,t)$ data (circles) for
  two packet size fractions: $\mu = 0.002$ (fast thermalization, blue
  circles) and $\mu = 0.8$ (slow thermalization, yellow squares). The
  continuous lines represent respectively an exponential decay fit
  for $\mu=0.8$, $C(t) = e^{-t/\tau_{\text{macro}}}$, with
  $\tau_{\text{macro}} = 672 \pm 5$, and an exponential plus stretched
  exponential for $\mu=0.002$,
  $C(t) = A(\tau_s)\,e^{-t/\tau_s} + B(\tau_c)\,e^{-\sqrt{t/\tau_c}}$,
  with $\tau_s = 1.45 \pm 0.09$ and $\tau_c = 141 \pm 12$.}
\label{fig:random-modes-panels}
\end{figure}

\subsection{Random modes packets}\label{subsec:random-modes}

In this section we present new results for the random modes, whose behaviour was studied in Ref.~\cite{COCCIAGLIA2022127581} for the harmonic chain. First of all, recall that each random mode is a random superposition of all Fourier modes. By construction, therefore, it has no intrinsic characteristic frequency, since all frequencies contribute to it, and its index $k$, unlike in the case of Fourier modes, is merely a label. In this respect, one of the main observations of Ref.~\cite{COCCIAGLIA2022127581} is that the time autocorrelation function of the energies of the random modes does not depend on the label $k$. The function
\begin{align}
  C_k^{(\rm R)}(\boldsymbol{\theta},t) = \frac{\left\langle \mathcal{E}^{(\rm R)}_k(\boldsymbol{\theta},t) \mathcal{E}^{(\rm R)}_k(\boldsymbol{\theta},0) \right\rangle -
    \left\langle \mathcal{E}^{(\rm R)}_k(\boldsymbol\theta) \right\rangle^2}{\left\langle \bigl[ \mathcal{E}^{(\rm R)}_k(\boldsymbol\theta) \bigr]^2\right\rangle - \left\langle \mathcal{E}^{(\rm R)}_k(\boldsymbol\theta) \right\rangle^2},
\label{eq:Ck-single-random}
\end{align}
decays exponentially with the same characteristic time for all values of $k$ (Fig.~\ref{fig:correlations-random-modes-single}). In Appendix \ref{app:neff_rotated} we provide an independent relaxation diagnostic based on spectral entropy. 

Since random modes, unlike Fourier modes, are not associated with a characteristic time scale, there is no difference between random and ordered packets. We therefore refer simply to packets of random modes of size $M$, without specifying how the modes are chosen. Figs.~\ref{fig:correlations-random-modes} and \ref{fig:autocorr_Ebar_rotated_fit} show the decay of the correlation functions for increasing packet size. Comparing Fig.~\ref{fig:correlations-random-modes} with Fig.~\ref{fig:autocorrelation-ordered-fourier-fits} reveals a remarkable difference between packets of Fourier modes and packets of random modes: for random modes, \emph{the larger the packet, the slower the decay of the energy autocorrelation}, which nevertheless remains exponential for all packet sizes. We conjecture that this behavior arises because random modes, unlike Fourier modes, are closer in nature to quasi-particle variables. Then, the larger the number of quasi-particles in the packet, the longer its relaxation time, since larger packets are more sensitive to the underlying integrability of the model.

In particular, Fig.~\ref{fig:autocorr_Ebar_rotated_fit} shows that, for small packets of random modes, several characteristic time scales clearly contribute to the relaxation: for a packet size ratio $\mu = 0.002$, the numerical data are well fitted by an exponential
plus a stretched exponential, whereas for a large ratio such as $\mu = 0.8$ they are well fitted by a simple exponential. In practice, for packets of random modes of size $M$ we find
\begin{align}
  \mu = M/N \ll 1 \qquad\rightarrow&\qquad C_{\boldsymbol{\theta}}^{(\rm R)}(M,t) \simeq A(\tau_s)\,e^{-t/\tau_s} + B(\tau_c)\,e^{-\sqrt{t/\tau_c}}, \nonumber \\
  \mu = M/N \sim 1 \qquad\rightarrow&\qquad C_{\boldsymbol{\theta}}^{(\rm R)}(M,t) \simeq e^{-t/\tau_{\text{macro}}}. 
\end{align}
Recall that a stretched exponential can be interpreted as the superposition of many exponential relaxations with different time scales. Indeed, even without a detailed analytical description of the relaxation mechanism, the following simple argument can be made. The Toda chain is integrable, and its motion on the invariant tori is characterized by $N$ frequencies, i.e., by $N$ characteristic time scales $\tau_k^{\text{Toda}}$. Determining the full spectrum of these time scales is beyond the scope of the present analysis: it would require diagonalizing the Lax matrix exactly and determining precisely how each time scale contributes to the relaxation of the random modes. Nevertheless, it is reasonable to assume that all Toda modes contribute to the relaxation of a packet of a few random modes, so that the correlation function can plausibly be written as
\begin{align}
  C_{\boldsymbol{\theta}}^{(\rm R)}(M,t) \simeq \sum_{k=1}^N p_\mu(\tau^{\rm Toda}_k)\,e^{-t/\tau^{\rm Toda}_k},
  \label{eq:C-random-modes}
\end{align}
where $p_\mu(\tau_k)$ is a probability distribution expressing our ignorance about the contribution of each Toda mode, depending on the relative size $\mu=M/N$ of the packet. If a single time scale $\tau_s=\tau^{\rm Toda}_{k^*}$ is well separated from the others, which instead form a quasi-continuum, Eq.~\eqref{eq:C-random-modes} can be written as
\begin{align}
  C_{\boldsymbol{\theta}}^{(\rm R)}(M,t) \simeq A(\tau_s)\,e^{-t/\tau_s} + B(\tau_c)\int_0^\infty d\tau\,p_\mu(\tau)\,e^{-t/\tau}.
  \label{eq:C-decomposed}
\end{align}
If, for instance,
\begin{equation}
  p_\mu(\tau) = \frac{\tau^{3/2}}{\sqrt{4\pi\tau_c}}\,e^{-\tau / 4\tau_c},
\end{equation}
the integral in Eq.~\eqref{eq:C-decomposed} yields exactly the stretched exponential $e^{-\sqrt{t/\tau_c}}$ that we fitted to the data for small $M$. For large $M$, instead, the data are well fitted by a single exponential with a long characteristic time. A plausible explanation is that the dynamics of a large packet of random modes is controlled by a collective mode governed by the slowest time scale of the model, which dominates over all the others. As we show in the next section, a very similar phenomenon occurs for the quasi-particle variables $\zeta_i$, introduced here for the first time.\\

\subsection{Quasi-particle modes packets}\label{subsec:quasiparticle-modes}

This section presents our results on the quasi-particle modes $\zeta_i$. As detailed in Appendix~\ref{app:zeta_loc}, the $\zeta_i$ are complex variables that receive contributions from all the $q_j$ and all the $p_j$, reading as
\begin{align}
\zeta_i = \sum_{j=1}^N \left[A^{(p)}_{ij} p_j + {\rm i} A^{(q)}_{ij} q_j\right],
\end{align}
where the precise definition of matrix elements for $A_{ij}^{(p)}$ and $A_{ij}^{(q)}$ is reported in App~\ref{app:zeta_loc}.
The dependence of $\zeta_i$ on the momenta is not localized, since the matrix elements $A^{(p)}_{ij}$ decay only as $A^{(p)}_{ij} \sim |i-j|^{-1/2}$ away from the diagonal. By contrast, its dependence on the positions is well resolved in space, since $A^{(q)}_{ij} \sim |i-j|^{-3/2}$. Because of this spatial resolution, we call the variables $\zeta_i$ quasi-particle modes, and we regard the proximity of their indices as an indication of contiguity in space. Accordingly, it makes sense to distinguish ordered packets, composed of neighboring quasi-particle modes, from random packets, composed of quasi-particle modes whose indices are drawn at random from the whole chain.

\begin{figure}[t!]
\centering
\begin{subfigure}[t]{0.49\textwidth}
    \centering
    \includegraphics[width=\textwidth]{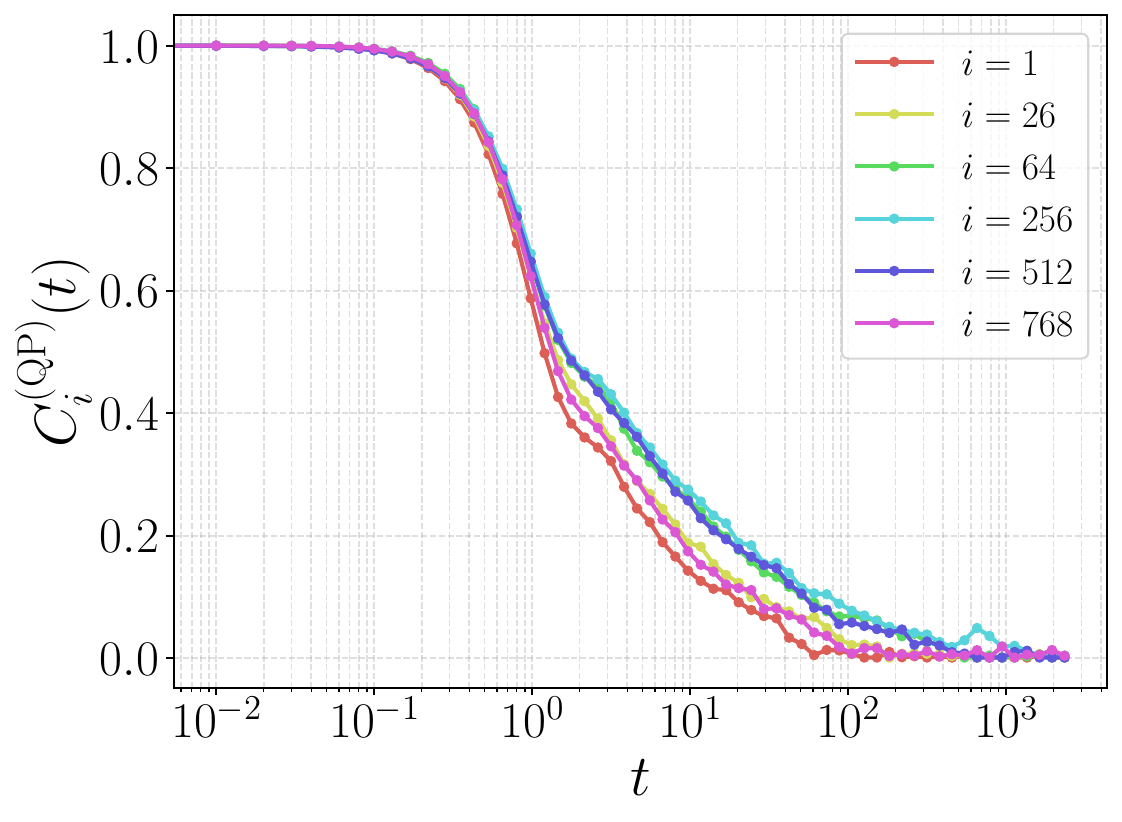}
    \caption{}
    \label{fig:zeta_singlemode_autocorr}
\end{subfigure}
\hfill
\begin{subfigure}[t]{0.49\textwidth}
    \centering
    \includegraphics[width=\textwidth]{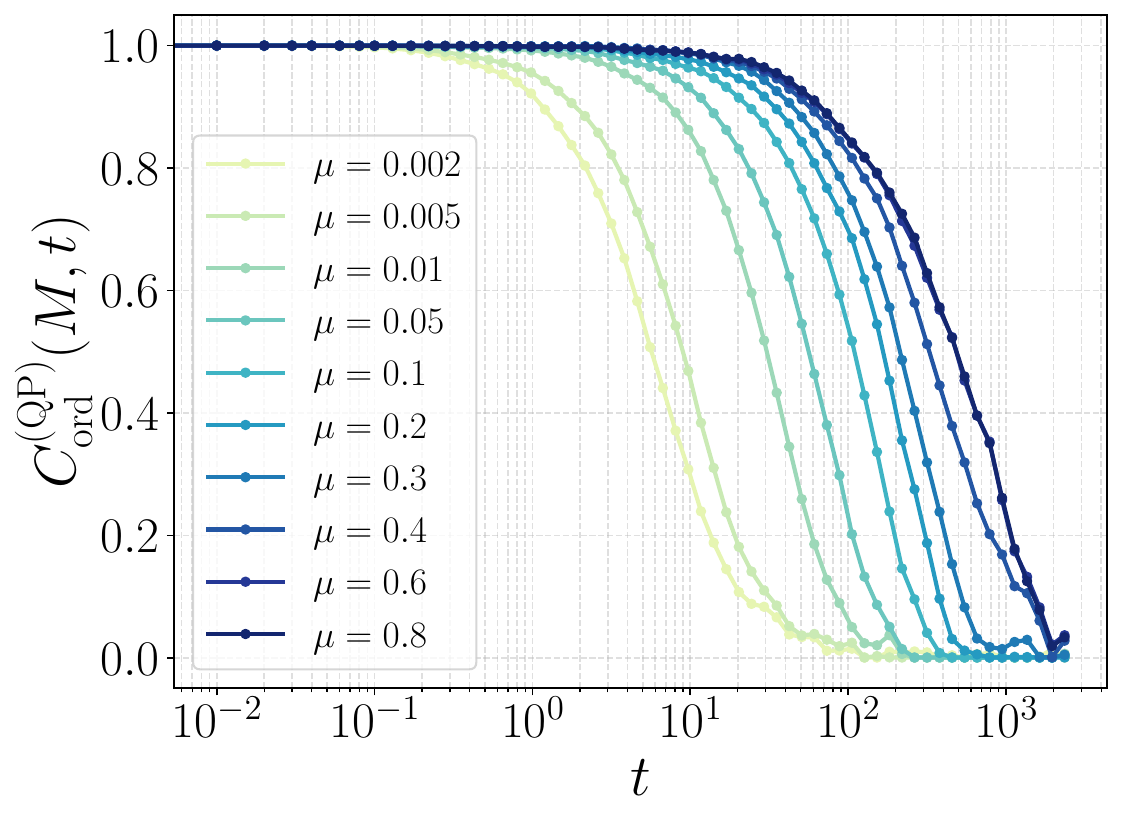}
    \caption{}
    \label{fig:zeta_ordered_packets}
\end{subfigure}
\hfill
\begin{subfigure}[t]{0.49\textwidth}
    \centering
    \includegraphics[width=\textwidth]{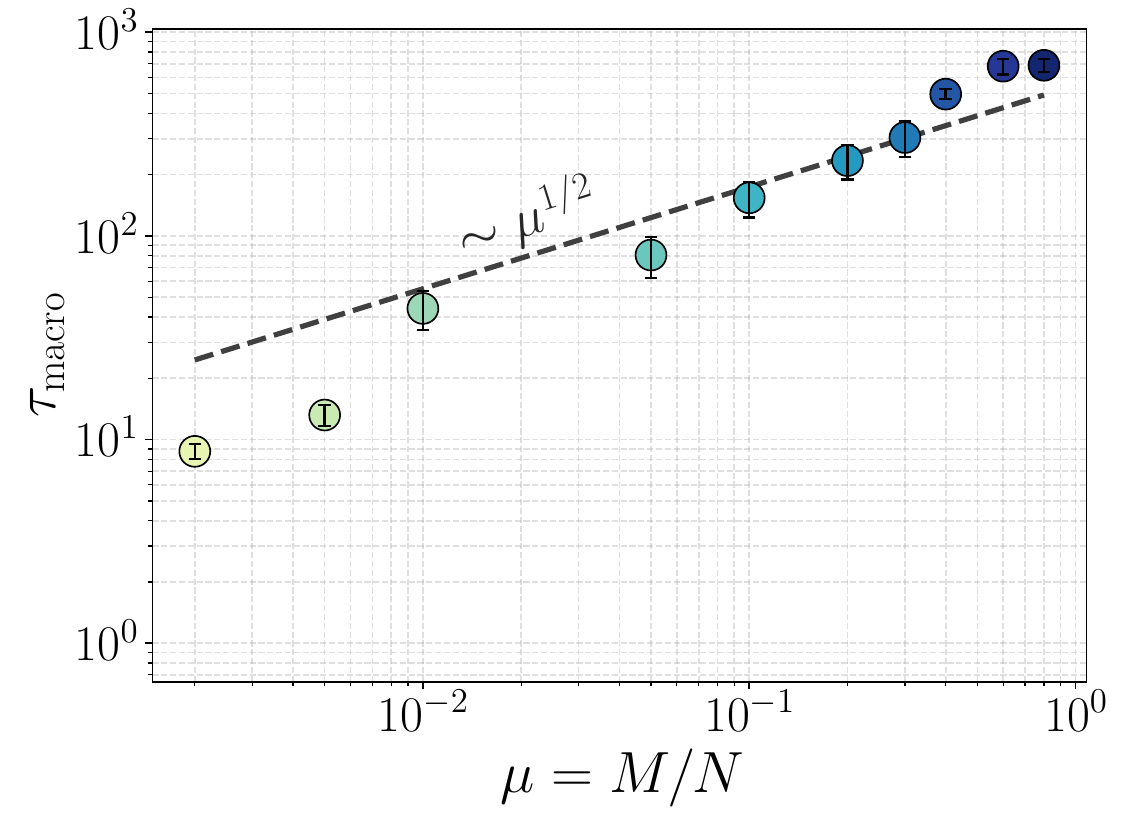}
    \caption{}
    \label{fig:zeta_tau_vs_M}
\end{subfigure}
\hfill
\caption{Time-correlation function of quasi-particle modes and mode packets. The Toda lattice has
  $N=1023$ particles and fixed boundary conditions. The initial condition of Hamiltonian dynamics is
  drawn from the thermal ensemble at $T=3$ ($\langle E \rangle/N =
  2.73$). Time step of Hamiltonian dynamics (leap-from symplectic
  integrator) is $\delta\tau=0.01$. \\ {\it \underline{Panel} a)}:
  Autocorrelation $C_i^{(\text{QP})}(t)$ of the single quasi-particle
  mode energies, see definition in
  Eq.~\eqref{eq:C-single-quasi-particle}; {\it \underline{Panel} b)}:
  Autocorrelation $C^{(\text{QP})}_\mathrm{ord}(M,t)$ of ordered
  quasi-particle packets, see definition in Eq.~\eqref{eq:CQP_ord},
  for packet fractions $\mu = M/N$ from $\mu=0.002$ to $\mu=0.8$. {\it
    \underline{Panel} c)}: Relaxation time $\tau_{\text{macro}}(\mu)$
  of the ordered quasi-particles packets (circles), obtained from
  exponential fit of data,
  $C^{(\text{QP})}_\mathrm{ord}(M,t)=e^{-t/\tau_{\text{macro}}}$, as a
  function of the packet size ratio $\mu=M/N$, compared with the power
  law $\tau_{\text{macro}}(\mu) \sim \sqrt{\tau_{\text{macro}}}$
  (dotted line).}
\label{fig:quasi-particles-panels-1}
\end{figure}

As in the previous sections, our first observation is that all quasi-particle modes individually relax exponentially to thermal equilibrium, with the same characteristic time scale regardless of the index $i$. This is shown in Fig.~\ref{fig:zeta_singlemode_autocorr}, which displays, for different values of $i$, the decay of the single quasi-particle energy autocorrelation
\begin{align}
  C_i^{(\rm QP)}(t) = \frac{\left\langle
    \mathcal{E}_i^{(\text{QP})}(t)
    \mathcal{E}_i^{(\text{QP})}(0)\right\rangle - \left\langle
    \mathcal{E}_i^{(\text{QP})}\right\rangle^2}{\left\langle
    \bigl[\mathcal{E}_i^{(\text{QP})}\bigr]^2\right\rangle - \left\langle
    \mathcal{E}_i^{(\text{QP})}\right\rangle^2}.
  \label{eq:C-single-quasi-particle}
\end{align}

\begin{figure}[t!]
\hfill
\begin{subfigure}[t]{0.49\textwidth}
    \centering
    \includegraphics[width=\textwidth]{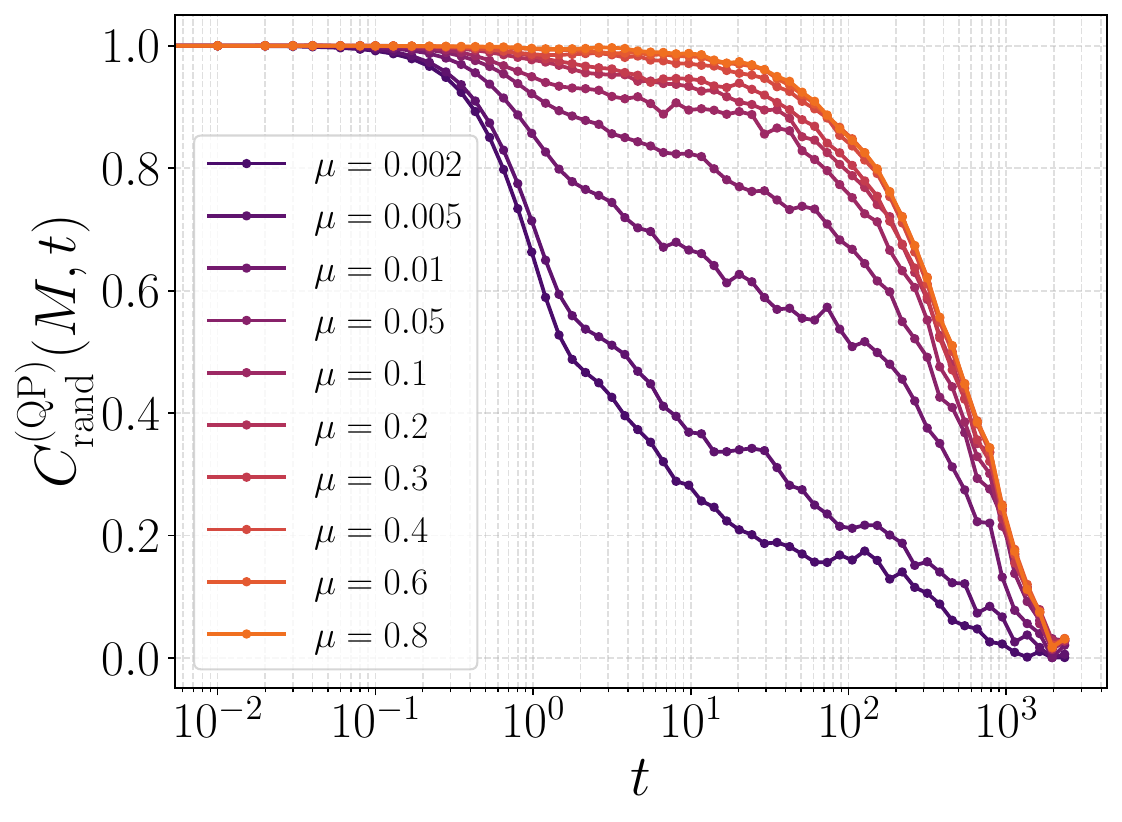}
    \caption{}
    \label{fig:zeta_random_packets}
\end{subfigure}
\hfill
\begin{subfigure}[t]{0.49\textwidth}
    \centering
    \includegraphics[width=\textwidth]{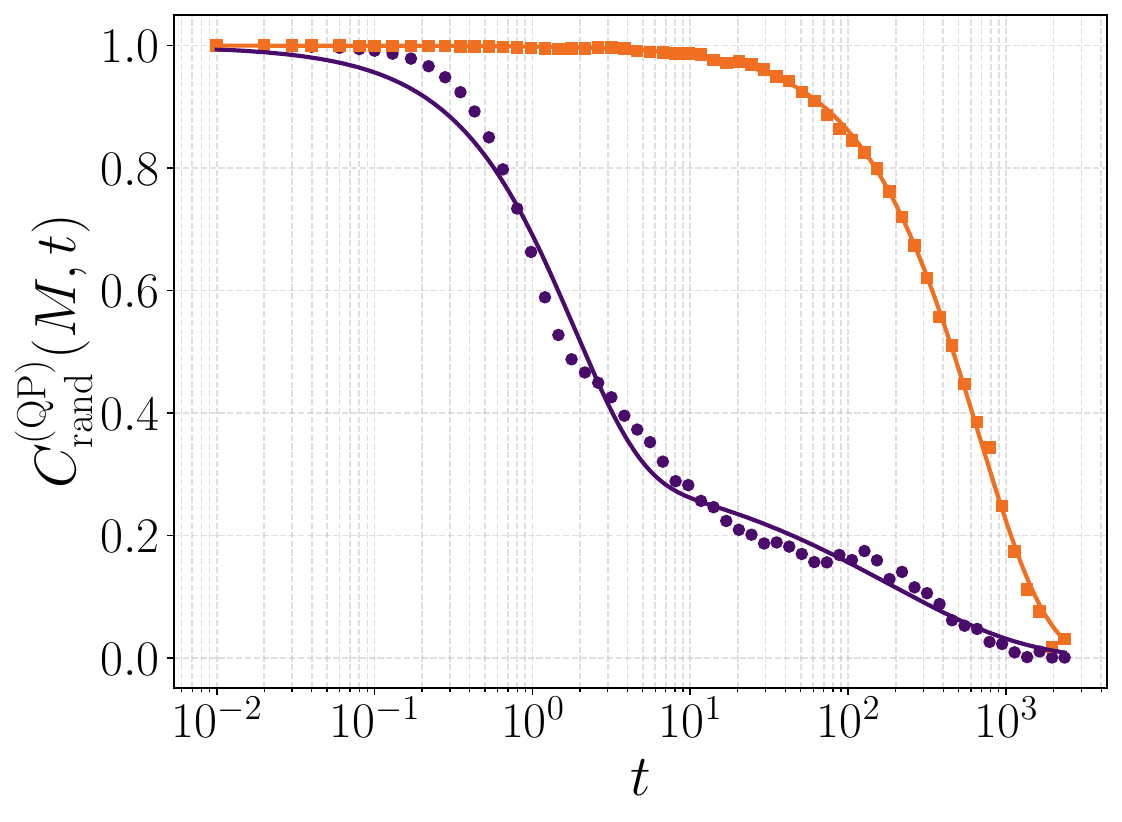}
    \caption{}
    \label{fig:zeta_random_packets_fits}
\end{subfigure}
\caption{Time correlation functions of quasi-particle {\it random} packets of
  modes in a Toda lattice with $N=1023$ particles and fixed boundary
  conditions. The initial condition of Hamiltonian dynamics is drawn
  from the thermal ensemble at $T=3$ ($\langle E \rangle/N =
  2.73$). Time step of Hamiltonian dynamics (leap-from symplectic
  integrator) is $\delta\tau=0.01$. {\it \underline{Panel} a)}:
  Normalized autocorrelation $C^{(\text{QP})}_{\text{rand}}(M,t)$ of
  quasi-particles packets, defined in Eq.~\eqref{eq:QP_dis}, for
  packets size fractions $\mu = M/N$ from $\mu=0.002$ to
  $\mu=0.8$. {\it \underline{Panel} b)}: Exponential fit
  $e^{-t/\tau_{\text{macro}}}$ of $C^{(\text{QP})}_{\text{rand}}(M,t)$
  = for packet size ratio $\mu = M/N=0.8$, with characteristic time
  $\tau_{\text{macro}} = 667 \pm 5$, and
stretched exponential fit $ C^{(\text{QP})}_{\text{rand}}(M,t) =
A(\tau_s)~e^{-t/\tau_s} + B(\tau_c)~e^{-\sqrt{t/\tau_c}}$ for
packet size ratio $\mu = M/N=0.002$, with characteristic times
$\tau_s = 1.81 \pm 0.10$ and $\tau_c = 184 \pm 37$.}
\label{fig:quasi-particles-panels-2}
\end{figure}

The energy autocorrelation of packets of quasi-particle modes is shown in Fig.~\ref{fig:zeta_ordered_packets} and in Fig.~\ref{fig:quasi-particles-panels-2}. Specifically, Fig.~\ref{fig:zeta_ordered_packets} shows the correlation functions $C^{(\text{QP})}_{\text{ord}}(M,t)$ of ordered packets, while Fig.~\ref{fig:quasi-particles-panels-2} shows that of randomly chosen packets, $C^{(\text{QP})}_{\text{rand}}(M,t)$. Recall that ordered packets consist of quasi-particle modes that are contiguous in space, whereas in random packets the modes are sampled across the whole chain. Thus, an ordered packet with size ratio $\mu = M/N$ corresponds to a contiguous fraction $\mu$ of the lattice. Instead, for a random packet the size ratio $\mu$ cannot be related to a spatial extent, since its quasi-particles are scattered across the whole system. Fig.~\ref{fig:zeta_ordered_packets} shows that for all ordered packets the energy autocorrelation decays exponentially, with a characteristic time that increases with the packet size. The dependence of these characteristic times $\tau_{\text{macro}}$ on the size ratio $\mu$ is reported in Fig.~\ref{fig:zeta_tau_vs_M}, which highlights the behavior $\tau_{\text{macro}} \sim \mu^{1/2}$. In the case of random packets, Fig.~\ref{fig:zeta_random_packets} shows that, as for the packets of random modes discussed in Sec.~\ref{subsec:random-modes}, the larger the packet, the longer the relaxation time, while the relaxation remains exponential. Fig.~\ref{fig:zeta_random_packets_fits} then shows explicitly that, as for random modes, the relaxation of large packets is well described by a single exponential. For small packets, instead, the data are well fitted by the superposition of a purely exponential decay at short times and a stretched exponential at later times, reproducing the behavior found for random modes:
\begin{align}
C^{(\text{QP})}_{\text{rand}}(M,t) = A(\tau_s)\,e^{-t/\tau_s} + B(\tau_c)\,e^{-\sqrt{\tau/\tau_c}}.
\end{align}
We stress that, also in this case, the exponent $1/2$ of the stretched exponential is not a fit parameter. It is fixed by the distribution of time scales $p_\mu(\tau)$ discussed in Sec.~\ref{subsec:random-modes}, and it agrees well with the numerical data.

\section{Conclusions}\label{sec:conclusions}

We have studied the relaxation to thermal equilibrium of different kinds of coarse-grained variables along the Hamiltonian flow of a classical integrable model, the one-dimensional Toda lattice. In particular, we have studied the relaxation of the energy fluctuations of packets of microscopic variables, namely Fourier modes, random modes and quasi-particle modes, with size ratios $\mu = M/N$ ranging from $\mu = 0.002$ to $\mu = 0.8$. In all cases, although the characteristic relaxation times depends on the size of the packet, and the dependence is different for different microscopic variables, the relaxation is always exponential or stretched exponential. Our findings support the viewpoint, advocated in a series of recent works~\cite{Baldovin_2021,COCCIAGLIA2022127581,Cattaneo25,Baldovin_review}, and inspired by Khinchin's approach to ergodicity, that an appropriate choice of observables, together with a large number of degrees of freedom, is sufficient to ensure the validity of statistical mechanics, i.e., the equivalence between dynamical and ensemble averages. Such an equivalence holds irrespective of the nature of the microscopic dynamics and, in particular, of the presence of chaos: this is clearly exemplified by our numerical results on the Toda lattice, in which, despite model's integrability, we have individuated several kind of coarse-grained variables which present good thermalization properties.

\backmatter

\begin{appendices}

\section{Spatial resolution of quasi-particle modes}\label{app:zeta_loc}

In this appendix we consider the variables $\zeta_i$ introduced in Sec.~\ref{subsec:into_quasi_particle}, in order to show that they describe excitations well resolved in real space. Let us start by writing explicitly the definition of $\zeta_i$ as
\begin{align}
    \zeta_i &= \sqrt{\frac{2}{N+1}}\sum_{k=1}^N \sin \left(\frac{\pi i k}{N+1}\right)z_k \nonumber\\
    &= \frac{2}{N+1}\sum_{j=1}^N \sum_{k=1}^N \sin \left(\frac{\pi i k}{N+1}\right)\sin\left(\frac{\pi j k}{N+1}\right)\left[\frac{p_j}{\sqrt{2\omega_k}}+\iunit\sqrt{\frac{\omega_k}{2}}\,q_j\right]\nonumber\\
    &= \sum_{j=1}^N \left[A^{(p)}_{ij}\, p_j + \iunit\, A^{(q)}_{ij}\, q_j\right],
\end{align}
where
\begin{align}
    A_{ij}^{(p)} &= \frac{\sqrt 2}{N+1}\sum_{k=1}^N \sin \left(\frac{\pi i k}{N+1}\right) \sin\left(\frac{\pi j k}{N+1}\right)\omega_k^{-1/2},\\
    A_{ij}^{(q)} &= \frac{\sqrt 2}{N+1}\sum_{k=1}^N \sin \left(\frac{\pi i k}{N+1}\right) \sin\left(\frac{\pi j k}{N+1}\right)\omega_k^{1/2}.
\end{align}
To understand the localization properties of $\zeta_i$, we study how the matrix elements $A_{ij}^{(p)}$ and $A_{ij}^{(q)}$ decay away from the diagonal $i=j$. Using the product-to-sum identity for sines, we write
\begin{equation}
    A_{ij}^{(p)} = \frac{1}{\sqrt 2(N+1)}\sum_{k=1}^N \left[\cos\left(\frac{\pi(i-j) k}{N+1}\right)-\cos\left(\frac{\pi (i+j) k}{N+1}\right)\right]\omega_k^{-1/2}.
\end{equation}
In the large-$N$ limit, introducing $\theta = \pi k/(N+1) \in (0,\pi)$ and using Eq.~\eqref{eq:dispersion}, this becomes
\begin{align}
    A_{ij}^{(p)} &\simeq \frac{1}{2}\int_0^\pi \frac{d\theta}{\pi}\left\{\cos[(i-j)\theta]-\cos[(i+j)\theta]\right\}\sin^{-1/2}(\theta/2)\nonumber\\
    &=\frac{I(i-j)-I(i+j)}{2},
\end{align}
where $I(m) = \int_0^\pi \frac{d\theta}{\pi}\cos(m\theta)\sin^{-1/2}(\theta/2)$. For large $m$, the leading contribution to this oscillatory integral comes from the lower endpoint $\theta=0$, where the amplitude $\sin^{-1/2}(\theta/2) \simeq \sqrt 2\,\theta^{-1/2} + \mathcal O(\theta^{3/2})$ has an integrable square-root singularity. In this limit, $I(m) \sim \int_0^\infty \frac{d\theta}{\pi} \cos(m\theta)\,\theta^{-1/2} \sim |m|^{-1/2}$. For sites far from the boundaries, the term $I(i+j)$ is subleading, and the matrix elements $A_{ij}^{(p)}$ decay as the power law
\begin{equation}
    A_{ij}^{(p)} \sim \frac{1}{|i-j|^{1/2}}
\end{equation}
away from the diagonal $i=j$.

A similar argument applies to $A_{ij}^{(q)}$. In this case
\begin{align}
    A_{ij}^{(q)} &= \frac{1}{\sqrt 2(N+1)}\sum_{k=1}^N \left[\cos\left(\frac{\pi(i-j) k}{N+1}\right)-\cos\left(\frac{\pi (i+j) k}{N+1}\right)\right]\omega_k^{1/2}\nonumber\\
    &\simeq \int_0^\pi \frac{d\theta}{\pi} \left\{\cos[(i-j)\theta]-\cos[(i+j)\theta]\right\}\sin^{1/2}(\theta/2)\nonumber\\
    &= J(i-j)-J(i+j),
\end{align}
where
$J(m) = \int_0^\pi \frac{d\theta}{\pi}\cos(m\theta)\sin^{1/2}(\theta/2)$.
Integrating by parts, we obtain
\begin{equation}
    J(m) = \frac{1}{\pi m}\Bigl[\sin(m\theta)\sin^{1/2}(\theta/2)\Bigr]^\pi_0
    - \frac{1}{4m}\int_0^\pi \frac{d\theta}{\pi}\, \sin(m\theta)\cos(\theta/2) \sin^{-1/2}(\theta/2).
\end{equation}
Since $m$ is an integer, the boundary term vanishes identically. The remaining oscillatory integral has amplitude $\cos(\theta/2) \sin^{-1/2}(\theta/2)\simeq \sqrt 2\,\theta^{-1/2} + \mathcal O(\theta^{3/2})$, with the same endpoint singularity as in $I(m)$, and therefore also scales as $|m|^{-1/2}$. Taking into account the prefactor $1/m$, it foloows that the matrix elements $A_{ij}^{(q)}$ decay as the power law
\begin{equation}
    A_{ij}^{(q)} \sim \frac{1}{|i-j|^{3/2}}
\end{equation}
away from the diagonal $i=j$.

We conclude that the $\zeta_i$ modes are not strictly localized, since the matrix elements $A_{ij}^{(p)}$ and $A_{ij}^{(q)}$ do not decay exponentially for $i\neq j$. They are, however, spatially resolved in their dependence on the positions: since $\sum_j |i-j|^{-3/2}$ converges, each $\zeta_i$ depends essentially on the displacements $q_j$ of the particles in a finite neighborhood of site $i$. By contrast, $\sum_j |i-j|^{-1/2}$ diverges, so the dependence on the momenta $p_j$ extends over the whole chain.

\section{Relaxation diagnostic of random modes via spectral entropy}\label{app:neff_rotated}

This appendix reports the non-equilibrium diagnostic used to verify that relaxation in the Haar-rotated representation is \emph{typical} with respect to the choice of the random rotation. We consider the Toda chain with fixed boundary conditions $q_0=q_{N+1}=0$ and integrate the dynamics in real space, $(q_i(t),p_i(t))_{i=1}^N$. The mode variables are defined by the sine transform~\eqref{eq:FourierModes}, from which we construct the complex Fourier amplitudes $z_k(t)=(P_k+\iunit\,\omega_k Q_k)/\sqrt{2\omega_k}$ and their Haar-rotated counterparts, i.e., the random modes $z_k(\boldsymbol{\theta},t)=\sum_{q=1}^N M_{kq}(\boldsymbol{\theta})\,z_q(t)$, with $M(\boldsymbol{\theta})\in U(N)$ Haar-distributed.

For each independent Haar realization $M(\boldsymbol{\theta})$, we prepare an out-of-equilibrium initial condition by exciting a \emph{single} rotated mode. In terms of the rotated complex amplitudes, we set
\begin{equation}
  z_k(\boldsymbol{\theta},0)=0 \quad \text{for } k\neq 1,
  \qquad
  z_1(\boldsymbol{\theta},0)\ \text{such that}\ 
  \mathcal{E}^{(\mathrm R)}_1(\boldsymbol\theta, 0)=E_0,
\end{equation}
where the energy of the $k$-th rotated mode is the diagonal term
\begin{equation}
  \mathcal{E}^{(\mathrm R)}_k(\boldsymbol\theta, t)=c_{kk}(\boldsymbol{\theta})\,|z_k(\boldsymbol{\theta},t)|^2 .
\end{equation}
Equivalently, the initial rotated-mode energies satisfy
\begin{equation}
  \mathcal{E}^{(\mathrm R)}_k(\boldsymbol \theta, 0)=
  \begin{cases}
    E_0, & k=1,\\[2mm]
    0, & k\neq 1.
  \end{cases}
  \label{eq:init-random}
\end{equation}
This initial state is mapped back to real space by inverting the rotation, the definition of $z_k$ and the sine transform, and the Toda equations of motion are then integrated. Since in the rotated variables the Hamiltonian contains both off-diagonal quadratic couplings and nonlinear terms, the diagonal energies $\mathcal{E}^{(\mathrm R)}_k(\boldsymbol\theta, t)$ are not conserved, and their redistribution provides a direct measure of relaxation in the rotated basis.

\begin{figure}[H]
\centering
\includegraphics[width=0.5\textwidth]{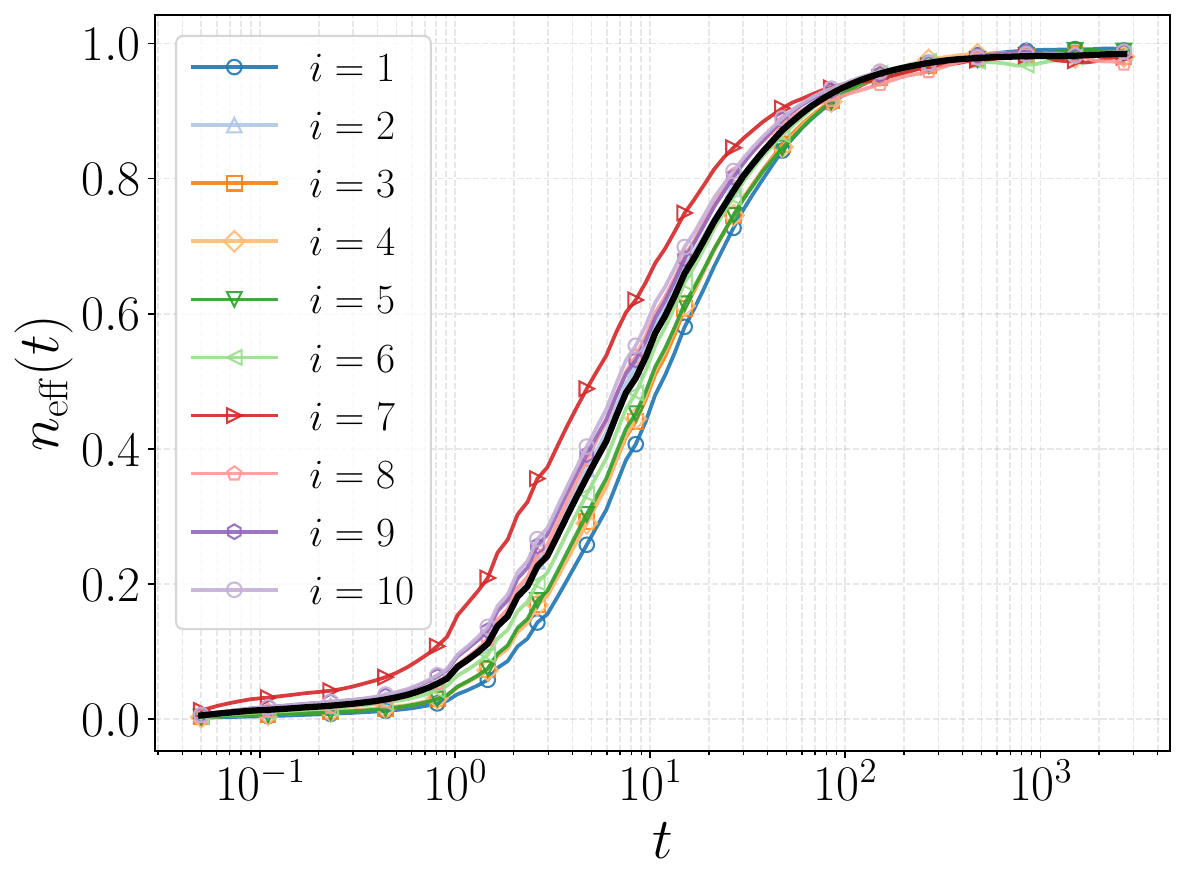}
\caption{Effective fraction of excited degrees of freedom $n_{\mathrm{eff}}(t)$ in the Haar-rotated basis for $N=1024$ under the single-mode injection protocol Eq.~\eqref{eq:init-random}. Thin colored curves: individual Haar realizations (run index $i$). Thick black curve: ensemble average $\overline{n_{\mathrm{eff}}}(t)$ over $50$ realizations.}
\label{fig:neff_rotated}
\end{figure}

We quantify the spreading of energy over the rotated labels through the normalized time-averaged energy fractions
\begin{equation}
  u_k(\boldsymbol{\theta},t)=
  \frac{\langle \mathcal{E}^{(\mathrm R)}_k (\boldsymbol\theta) \rangle_t}
       {\sum_{q=1}^N \langle \mathcal{E}^{(\mathrm R)}_q(\boldsymbol\theta)\rangle_t},
  \qquad
  \langle \mathcal{E}^{(\mathrm R)}_k(\boldsymbol\theta)\rangle_t
  =\frac{1}{t}\int_0^t ds\,\mathcal{E}^{(\mathrm R)}_k(\boldsymbol\theta, s),
\end{equation}
the associated spectral entropy
\begin{equation}
  S_{\mathrm{sp}}(\boldsymbol{\theta},t)=-\sum_{k=1}^N u_k(\boldsymbol{\theta},t)\,\log u_k(\boldsymbol{\theta},t),
\end{equation}
and the effective fraction of excited degrees of freedom
\begin{equation}
  n_{\mathrm{eff}}(\boldsymbol{\theta},t)=\frac{\exp(S_{\mathrm{sp}}(\boldsymbol{\theta},t))}{N},
\end{equation}
which grows from $1/N$, for a single excited mode, to unity at equipartition of the time-averaged energies.  We repeat the protocol for $M_{\mathrm{run}}$ independent Haar realizations and compute $n_{\mathrm{eff}}^{(i)}(t)$ for each run $i$, together with the ensemble average 
\begin{equation}
\overline{n_{\mathrm{eff}}}(t)=\frac{1}{M_{\mathrm{run}}}\sum_{i=1}^{M_{\mathrm{run}}} n_{\mathrm{eff}}^{(i)}(t).
\end{equation}
Fig.~\ref{fig:neff_rotated} shows $n_{\mathrm{eff}}^{(i)}(t)$ for representative realizations (thin colored curves) and the average over all runs (thick black curve). The close collapse of the curves indicates that the relaxation profile of $n_{\mathrm{eff}}(t)$ is typical and only weakly dependent on the chosen Haar rotation. This robustness with respect to the basis motivates the use of Haar-rotated mode observables in the main text, where we focus instead on equilibrium time correlations of coarse-grained packet energies (Sec.~\ref{sec:autocorrelationcoarsegrained}).

\section{Monte Carlo initialization}\label{app:mc_ic}

Here we describe the procedure used to initialize the Toda chain for the results presented in this paper. To avoid any bias associated with special harmonic initial states, such as fixed phases, coherent patterns or even ad hoc random-phase choices, we generate initial conditions typical of the thermal Gibbs measure by means of Monte Carlo sampling. This matters for the questions addressed in this work, since packet relaxation and long-time memory can be strongly affected by subtle correlations introduced at $t=0$ by structured mode constructions. Our goal is therefore to start from an agnostic class of states that does not privilege any basis: we thermalize the canonical variables $(q_i,p_i)$ with respect to the full nonlinear Toda Hamiltonian.

We consider the Toda chain with fixed boundary conditions $q_0=q_{N+1}=0$ and Hamiltonian
\begin{equation}
  H(q,p)=\sum_{i=1}^N \frac{p_i^2}{2}+U(q),
  \qquad
  U(q)=\sum_{i=0}^{N}V(q_{i+1}-q_i),
  \label{eq:H_mc_appendix}
\end{equation}
where $V(x)=e^{-x}+x-1$ is the Toda potential. The boundary conditions enter $U(q)$ through the outermost differences $q_1-q_0=q_1$ and $q_{N+1}-q_N=-q_N$. At temperature $T$, the canonical Gibbs density is
\begin{equation}
  \rho_T(q,p)=\frac{1}{Z_T}\,e^{-\beta H(q,p)},\qquad \beta=\frac{1}{T},
\end{equation}
with partition function $Z_T=\int dq\,dp\,e^{-\beta H(q,p)}$. Since the kinetic energy is a sum of independent quadratic terms and $U$ depends only on $q$, the density factorizes exactly as
\begin{equation}
  \rho_T(q,p)=\rho_T^{(p)}(p)\,\rho_T^{(q)}(q),
  \qquad
  \rho_T^{(p)}(p)\propto e^{-\beta\sum_i p_i^2/2},
  \qquad
  \rho_T^{(q)}(q)\propto e^{-\beta U(q)}.
  \label{eq:gibbs_factorization_appendix}
\end{equation}
The momenta can therefore be sampled independently from the Maxwell distribution,
\begin{equation}
  p_i \sim \mathcal{N}(0,T),
  \qquad i=1,\dots,N,
  \label{eq:maxwell_momenta_appendix}
\end{equation}
while the nontrivial step is sampling the configurational Boltzmann weight $\rho_T^{(q)}(q)\propto e^{-\beta U(q)}$. To sample $q$ from $\rho_T^{(q)}$, we run a Metropolis-Hastings Markov chain with stationary measure
\begin{equation}
  \pi_T(q)\propto e^{-\beta U(q)}.
\end{equation}
Starting from the reference configuration $q_i=0$, each Monte Carlo move proposes a local update at a single, randomly chosen site $i$,
\begin{equation}
  q_i \to q_i' = q_i+\delta,
  \qquad
  \delta \sim \mathrm{Unif}[-\Delta,\Delta],
  \label{eq:local_proposal_appendix}
\end{equation}
leaving all other coordinates unchanged. The proposal is symmetric in $\delta$, detailed balance with respect to $\pi_T$ is ensured by the standard Metropolis acceptance probability
\begin{equation}
  \mathcal{P}(q\to q')=\min\!\Big\{1,\exp\!\big(-\beta\,\Delta U\big)\Big\},
  \qquad
  \Delta U = U(q')-U(q).
  \label{eq:metropolis_accept_appendix}
\end{equation}
One Monte Carlo sweep consists of $N$ such single-site proposals, with the updated site chosen uniformly at random each time. In our simulations we use $\Delta=4.0$, which yields an acceptance rate $\mathcal{P}(q\to q')\simeq 0.4$, and we perform of the order of $10^5$ Monte Carlo steps before starting the Hamiltonian dynamics.

\begin{figure}[H]
  \centering
  \begin{subfigure}{0.48\textwidth}
    \centering
    \includegraphics[width=\textwidth]{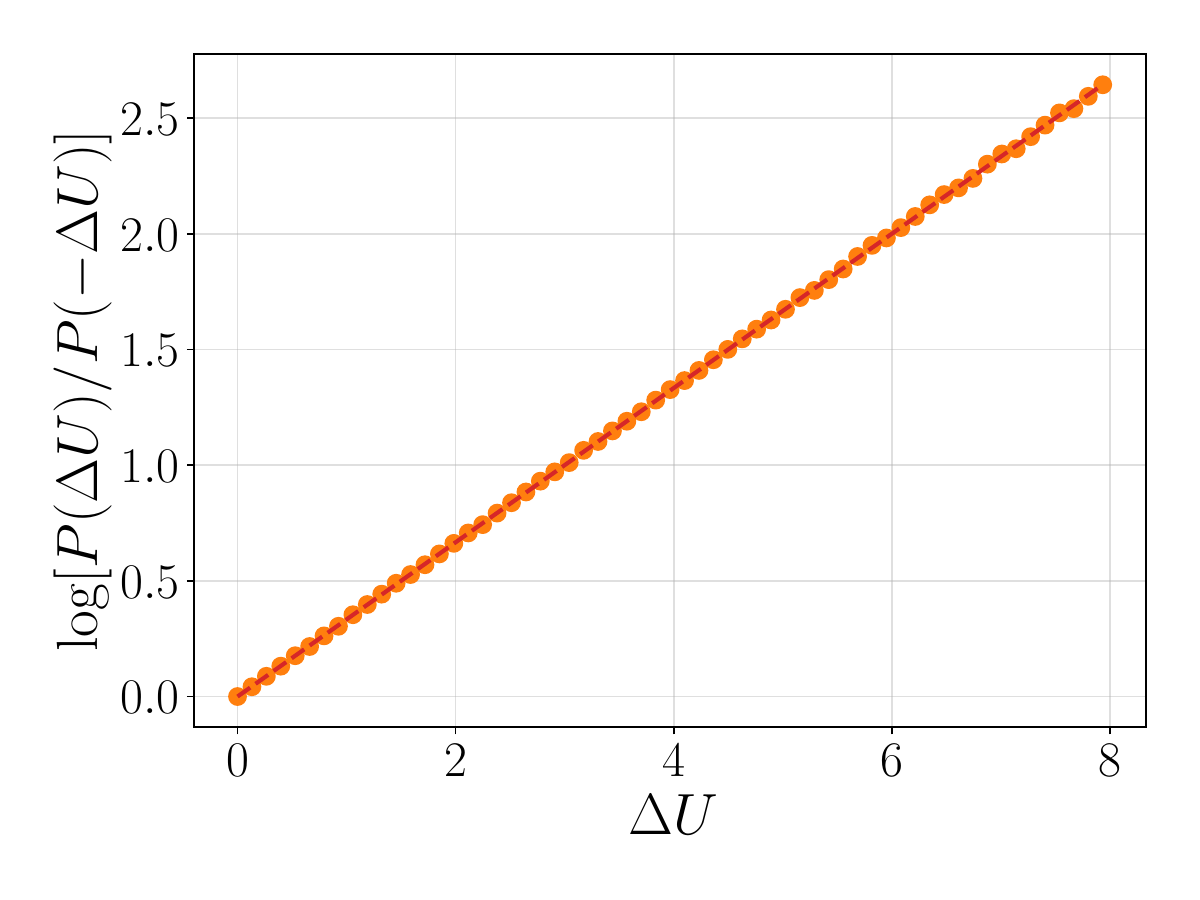}
    \caption{Log-ratio and linear fit}
    \label{fig:mc_boltzmann_test-1}
  \end{subfigure}
  \hfill
  \begin{subfigure}{0.48\textwidth}
    \centering
    \includegraphics[width=\textwidth]{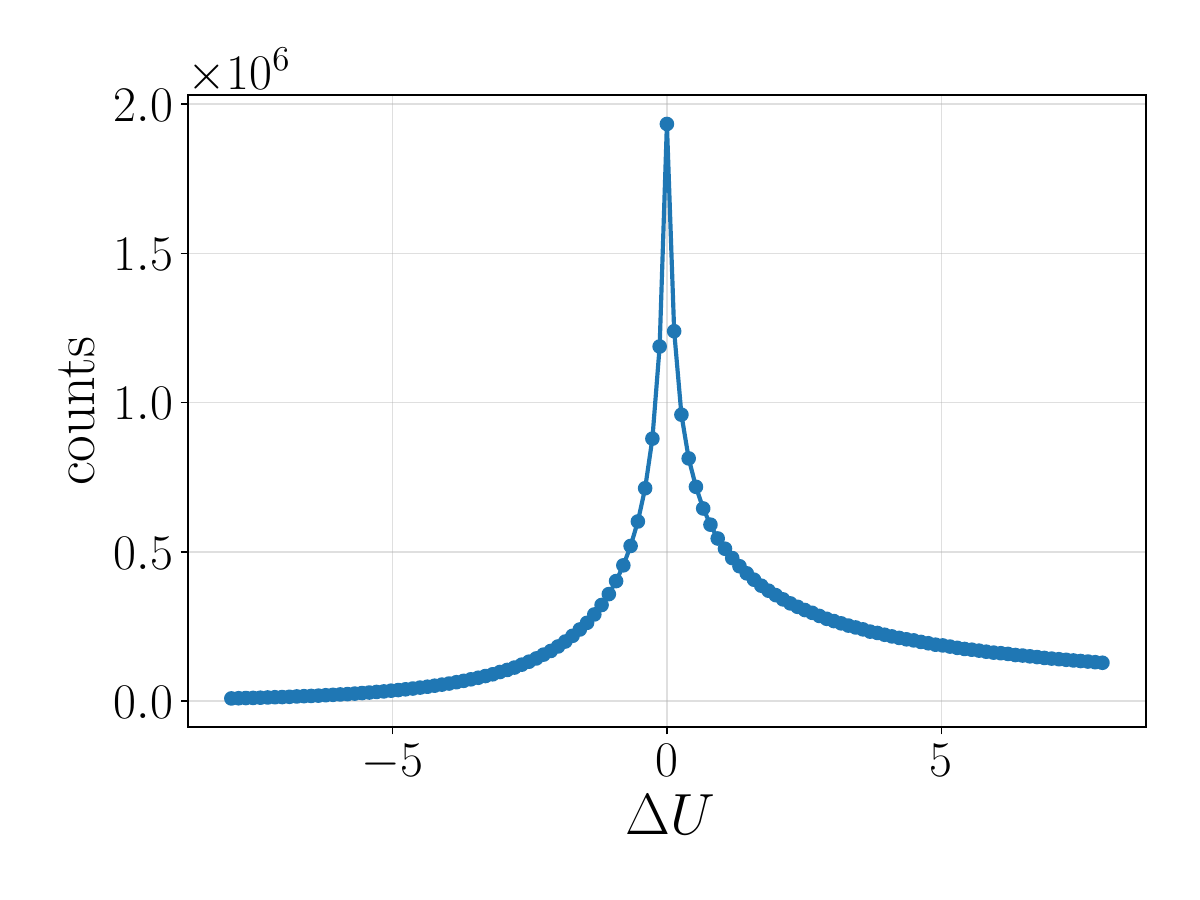}
    \caption{Histogram of $\Delta U$}
    \label{fig:mc_boltzmann_test-2}
  \end{subfigure}
  \caption{
  Boltzmann symmetry test for the proposed Monte Carlo energy increments. Left: measured quantity $\log[P(\Delta U)/P(-\Delta U)]$ as a function of $\Delta U$, together with a linear fit $b\,\Delta U$. The slope is found to be $b\simeq\beta=1/T$, in agreement with the canonical distribution. Right: histogram of proposed energy increments $\Delta U$, illustrating the asymmetry between positive and negative fluctuations consistent with the Boltzmann weight. The Toda Lattice for this simulation is made of $N=1023$ particles and total energy $E/N=2.73$. The temperature for the initial Monte Carlo procedure was set to $T=3$. Time step for integration $\Delta\tau=0.01$.
  }
  \label{fig:mc_boltzmann_test}
\end{figure}

\begin{figure}[H]
  \centering
  \begin{subfigure}{0.47\textwidth}
    \centering
    \includegraphics[width=\textwidth]{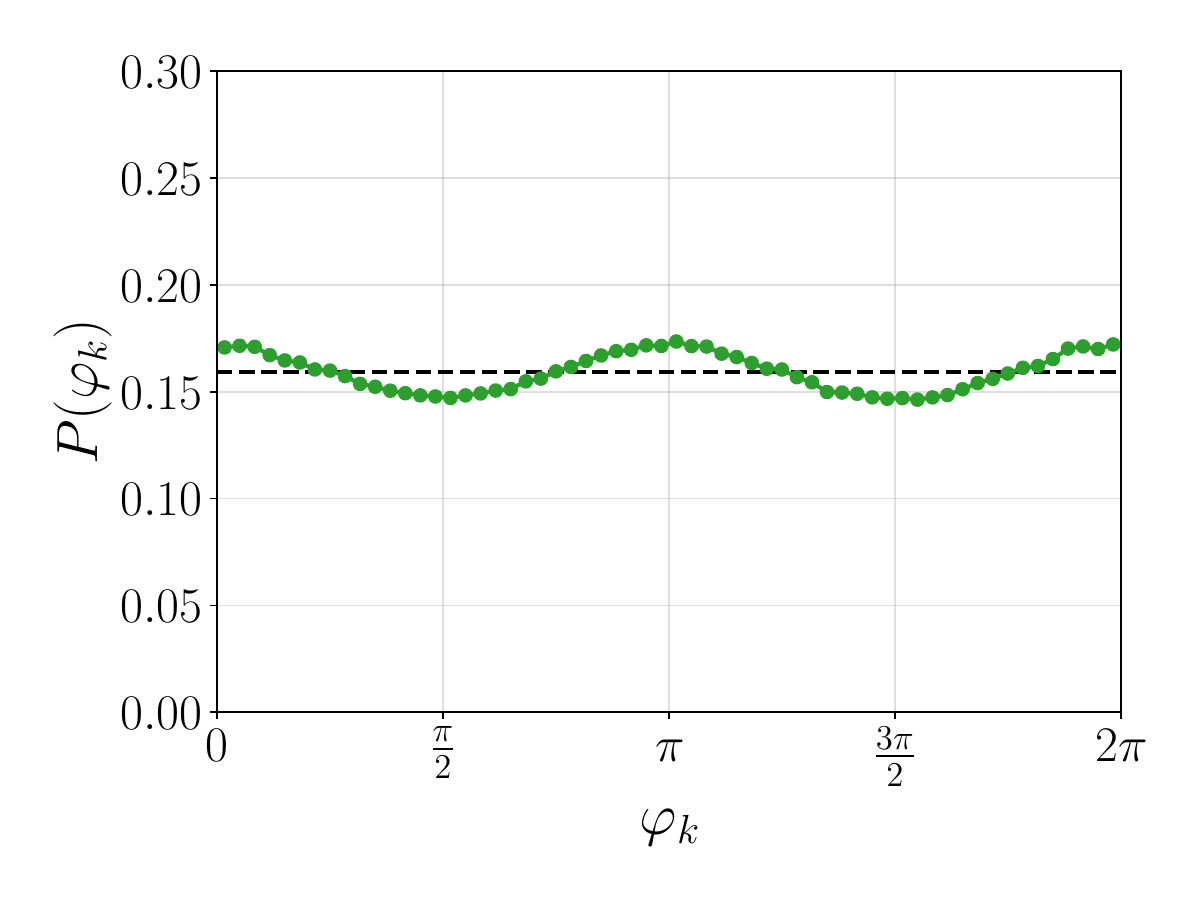}
    \caption{Phase distribution}
    \label{fig:phase_distribution}
  \end{subfigure}
  \begin{subfigure}{0.47\textwidth}
    \centering
    \includegraphics[width=\textwidth]{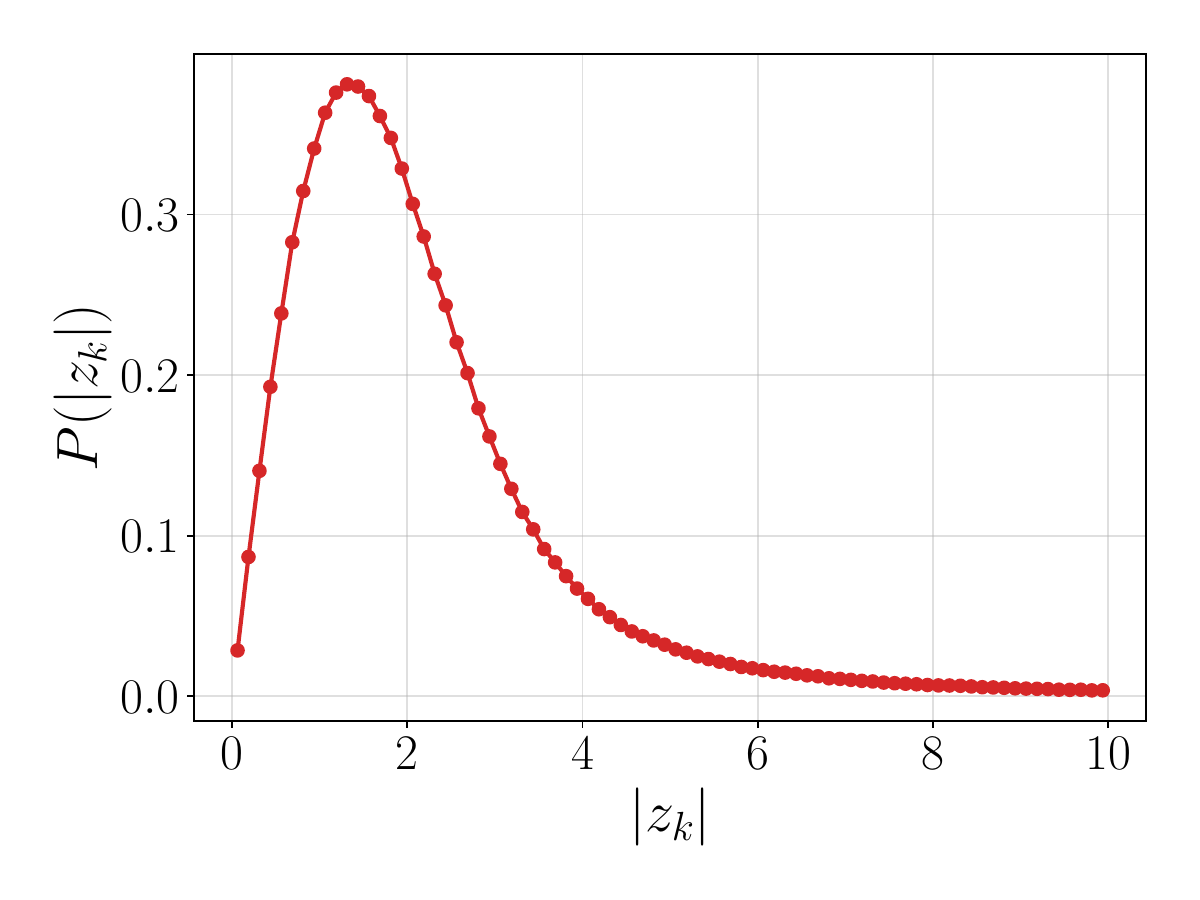}
    \caption{Modulus distribution}
    \label{fig:modulus_distribution}
  \end{subfigure}
  \caption{
  Left: probability distribution of the phases $\varphi_k$, compared with the uniform value $1/(2\pi)$. Right: probability distribution of the moduli $|z_k|$. The Toda Lattice for this simulation is made of $N=1023$ particles and total energy $E/N=2.73$. The temperature for the initial Monte Carlo procedure was set to $T=3$. Time step for integration $\Delta\tau=0.01$.
  }
  \label{fig:phase_modulus}
\end{figure}

As an additional equilibrium diagnostic, we test the Boltzmann symmetry of the proposed potential-energy increments. For symmetric single-site proposals and a stationary canonical measure $\pi_T$, the empirical distribution of proposed increments $\Delta U$ measured along the stationary Markov chain must satisfy
\begin{equation}
  \frac{P(\Delta U)}{P(-\Delta U)} = e^{\beta\,\Delta U},
\end{equation}
where $P(\Delta U)$ denotes the probability density of the proposed energy change $\Delta U = U_{\mathrm{after}}-U_{\mathrm{before}}$. In the simulation, we accumulate a histogram of $\Delta U$ over all proposed Monte Carlo moves (Fig.~\ref{fig:mc_boltzmann_test-2}), construct the logarithmic ratio $\log[P(\Delta U)/P(-\Delta U)]$, and fit it with a linear law $b\,\Delta U$. Consistency with the canonical Boltzmann weight therefore corresponds to fit parameter $b \simeq \beta = 1/T$. The result of this test is shown in Fig.~\ref{fig:mc_boltzmann_test}. The measured log-ratio is accurately linear in $\Delta U$, and the fitted slope is in excellent agreement with the imposed inverse temperature.
\noindent Furthermore, to ascertain that no residual coherent structures are present, we analyze the moduli and phases of the harmonic complex amplitudes
\begin{equation}
  z_k=\frac{P_k + {\rm i}\,\omega_k Q_k}{\sqrt{2\omega_k}},
  \qquad
  \varphi_k=\arg z_k\in[0,2\pi),
\end{equation}
where $(Q_k,P_k)$ are the sine-Fourier canonical coordinates and $\omega_k$ the corresponding linear frequencies. In thermal equilibrium, the phases $\varphi_k$ are expected to be approximately uniformly distributed over $[0,2\pi)$. Pronounced peaks or clustering in their histogram would instead signal residual phase correlations. As visible in Fig.~\ref{fig:phase_distribution}, the phase distribution is found to be close to uniform, indicating the absence of significant phase locking. If we now inspect the distribution of the moduli $|z_k|$ in Fig.~\ref{fig:modulus_distribution}, we observe that the resulting histogram  qualitatively resembles a Rayleigh distribution, consistent with a smooth, thermal-like statistics of the mode amplitudes.

\noindent Once the diagnostics of Figs.~\ref{fig:mc_boltzmann_test}-\ref{fig:phase_modulus} confirm adequate equilibration, the final Monte Carlo configuration $(q(0),p(0))$ is passed to the Hamiltonian integrator as the initial condition for the Toda equations of motion. From this point the thermostat is switched off and the subsequent evolution is purely deterministic and microcanonical at energy $E=H(q(0),p(0))$. This provides a clean baseline for the Hamiltonian evolution: any coherent features that appear later in Fourier, random-mode, or quasi-particle representations are generated dynamically rather than being planted in the initial condition by construction.

\end{appendices}

\bibliography{References}

\end{document}